\documentclass[twocolumn]{aastex631}

\usepackage{savesym}
\savesymbol{tablenum}
\usepackage{siunitx}
\usepackage{amsmath}
\usepackage{float}

\shortauthors{Hagey, Edwards \& Boley}
\graphicspath{{./}{figures/}}

\begin{document}
\title{Evidence of Long-Term Period Variations in the Exoplanet Transit Database (ETD)}
\shorttitle{Transit Variations in the ETD}

%%% AUTHORS %%%

\correspondingauthor{Simone Hagey}
\email{shagey@phas.ubc.ca}

\author[0000-0001-8072-0590]{Simone R. Hagey}
\affiliation{The University of British Columbia \\
6224 Agricultural Road Vancouver, BC V6T 1Z1, Canada \\}

\author[0000-0002-5494-3237]{Billy Edwards}
\affiliation{AIM, CEA, CNRS, Universit\'e Paris-Saclay, Universit\'e de Paris, F-91191 Gif-sur-Yvette, France}
\affiliation{Blue Skies Space Ltd., 69 Wilson Street, London, EC2A 2BB, UK}
\affiliation{Department of Physics and Astronomy, University College London, Gower Street, London, WC1E 6BT, UK}
\affiliation{Paris Region Fellow}

\author[0000-0002-0574-4418]{Aaron C. Boley}
\affiliation{The University of British Columbia \\
6224 Agricultural Road Vancouver, BC V6T 1Z1, Canada \\}
\affiliation{Canada Research Chair in Planetary Astronomy}

\submitjournal{The Astronomical Journal}
\accepted{September 9, 2022}

\begin{abstract}

We analyze a large number of citizen science data and identify eight Hot Jupiter systems that show evidence for deviations from a constant orbital period: HAT-P-19\,b, HAT-P-32\,b, TrES-1\,b, TrES-2\,b, TrES-5\,b, WASP-4\,b, WASP-10\,b, and WASP-12\,b. The latter system is already well known to exhibit strong evidence for tidal orbital decay and serves as an important control for this study. Several other systems we identify have disputed period drifts in the literature, allowing the results here to serve as an independent analysis. The citizen science data are from the Exoplanet Transit Database (ETD), which is a global project established in 2008 by the Variable Star and Exoplanet Section of the Czech Astronomical Society. With over 400 planets and 12,000 contributed observations spanning 15 years, the ETD is brimming with potential for studying the long-term orbital evolution of close-in Hot Jupiters. We use our results to discuss prioritization of targets for follow up investigations, which will be necessary to confirm the period drifts and their causes. 

\end{abstract}

\keywords{Exoplanets (498) --- Exoplanet dynamics (490) --- Transit timing variation method (1710)}

%%% INTRODUCTION %%%
\section{Introduction} \label{sec:intro}

Hot Jupiters (HJs) are gas giant planets that orbit their host stars with periods less than about 12 days. Their provenance remains debated, although several possible formation pathways exist, such as dynamical excitation followed by tidal circularization \citep{fabrycky_HJ_tides}, large scale disc migration \citep{lin_orbital_1996}, and in situ formation \citep{boley_situ_2016, batygin_situ_2016}. Regardless of how they formed, HJs' proximity to their host stars leads to tidal interactions, which in turn should affect their orbital evolution. Specifically, many HJ orbits should be shrinking gradually over time due to a transfer of angular momentum from the planet to the host star, eventually leading to planetary engulfment \citep{levrard_falling_2009,matsumura_tidal_2010}. This situation occurs when the orbital period of a HJ is shorter than the rotational period of its host star \citep{penev_empirical_2018} or if the angular momentum vectors are anti-aligned. For a given arrangement, the decay rate can vary depending on the magnitude of stellar tidal dissipation \citep{levrard_falling_2009,matsumura_tidal_2010}. Directly measuring the orbital decay rate provides an estimate of the stellar tidal quality factor $Q_*$. \par

While theory predicts many HJs should be decaying, direct detection has remained difficult. 
Nonetheless, \citet{patra_continuing_2020} suggest that certain ensemble properties of HJs indirectly reveal evidence of orbital decay -- properties such as the infrequency of planets with periods less than one day, the rapid rotation of some host stars, and the lack of short period planets around such stars.\par

In principle, careful observations of planetary transit centres over decade timescales should reveal direct evidence for orbital decay \citep{birkby_wts-2_2014}, appearing as a quadratic timing variation. However, orbital decay is not the only possible scenario for HJ orbital evolution, and some processes cause transit variations that mimic  decay. For instance, apsidal orbital precession causes a sinusoidal variation in transit times. Even for fast precession rates, predicted precession periods extend over many decades \citep{ragozzine_probing_2009}, making the sinusoidal behaviour very difficult to detect. Because of this, the curvature of the transit timing variations from apsidal precession can be consistent with the signal of orbital decay over (relatively) short observing windows. It must be stressed that measuring the apsidal precession rate of a HJ is valuable on its own, as it provides an estimate of the planetary Love number which enables probing the interior density distribution \citep{ragozzine_probing_2009}. Yet other effects may also cause observable transit time variations that do not reflect an inherent change in an HJ's orbit. A line-of-sight acceleration is one such example, in which a wide-orbit companion causes the host star and its HJ to accelerate toward the observer, leading to an apparent decay of the orbital period \citep{bouma_etal_2020}. Stellar activity may also lead to a perceived change in a planet’s orbital period on timescales comparable to current HJ observational baselines. A transit lightcurve is deformed if the planet passes over a starspot, which can affect the fitted transit centre time. In the case of persistent starspots, there could be a continuous perturbation of the fitted transit center times. \citet{ioannidis_p_how_2016} found that in rare cases this periodic nature could mimic TTVs induced by planetary companions, and \citet{patra_continuing_2020} suggest that stellar activity cycles with very long periods could mimic orbital decay. Investigating the effects of stellar activity and line-of-sight acceleration on the transit timing of the ETD targets is beyond the scope of this work, but must be considered in follow-up studies.\par

Whatever the cause, the earliest known HJs have transit observational baselines that span over a decade, making it possible to detect orbital decay or other long-term transit timing variations. Though many systems of interest have been suggested \citep[see][]{patra_continuing_2020}, only WASP-12\,b has a clear signature of spiraling into its host star. Indeed, since \citet{maciejewski_departure_2016} first suggested that WASP-12\,b does not follow a constant period ephemeris, the accumulation of more transit and secondary eclipse data has continuously supported an orbital decay model \citep{patra_apparently_2017,yee_orbit_2019,turner_decaying_2020}. With the inclusion of observations from the Transiting Exoplanet Survey Satellite \citep[TESS][]{ricker_transiting_2015}, the apsidal precession model has largely been rejected as well \citep{turner_decaying_2020}. This work has sparked increased interest within the exoplanet community in searching for decaying HJs, but as of now no other systems have shown such compelling results.\par

Given the large number of transiting planets that have been discovered, regularly obtaining follow-up observations for each of them has become a difficult task given the limited resources of professional facilities and the perceived low science yield of such observations. However, given their large transit depth, high quality observations of HJs around bright stars can be achieved with relatively modest equipment. As such, citizen astronomers have for decades been able to observe exoplanetary transits; and importantly, projects such as the Exoplanet Transit Database (ETD), ExoClock \citep{kokori_2021,kokori_2022} and Exoplanet Watch \citep{zellem_2020} provide platforms that allow these observers to collectively contribute to cataloguing light curves.\par

Amateur exoplanet transit observations further contribute to updating ephemerides and maximizing observing efficiency at major facilities by ensuring reliable timing. The TESS mission, in particular, has developed the TESS Follow-Up Observing Program (TFOP) to enlist amateur observers to aid in eliminating false-positives from TESS exoplanet candidates \citep{collins_tess_2019}. Furthermore, via the ORBYTS programme \citep{sousa_silva_orbyts} high-school students have contributed to refining the ephemerides of potential targets for the ESA Ariel mission \citep{tinetti_ariel,tinetti_ariel2} using the Las Cumbres Observatory Global Telescope (LCOGT) network of robotic telescopes \citep[e.g.][]{edwards_orbytsI,edwards_orbytsII}. There are also studies in the literature that use select citizen science transits for transit timing variation (TTV) studies \citep[see e.g.][]{baluev_benchmarking_2015,petrucci_discarding_2019,sonbas_probing_2021}. Nonetheless, the exclusive use of such amateur observations are rare, despite their potential for examining long-term orbital changes such as period decay. In this work, we analyze the wealth of citizen scientist observations spanning over a decade in the ETD to demonstrate these data sets can be used to identify HJ systems that are strong candidates for exhibiting observable orbital evolution. Such selections can then be followed-up using large facilities, making efficient use of resources.\par

In Section \ref{sec:data}, we describe the ETD, and in Section \ref{sec:targetselection} outline our target selection process. Section \ref{sec:methods} introduces the transit timing models and model fitting processes. Our results are presented in Section \ref{sec:results} and discussed in detail, with an emphasis on prospects for future observations, in Section \ref{sec:discussion}. This work would not be possible without the hundreds of citizen scientists who submitted transit observations to the ETD and we ask the reader to take note of the acknowledgements at the end of this paper.\par

%%% DATA %%%

\section{DATA} \label{sec:data}

The data used in this study are from the Exoplanet Transit Database (ETD). The ETD was established in 2008 by the Variable Star and Exoplanet Section of the Czech Astronomical Society, and allows any observer to register and upload transit observations \citep{poddany_exoplanet_2010,poddany_new_2011}. As of early 2022, there are 399 planets listed in the database and over 12,000 contributed observations.\footnote{\url{http://var2.astro.cz/ETD/index.php}} The website also provides transit predictions and a lightcurve fitting tool, requiring users to upload their target flux, error, and timestamps. The user is able to upload data with either a geocentric or heliocentric JD format based on the UTC time standard. The fitted transit times are then provided by the database in HJD$_{\rm UTC}$, which we convert to BJD$_{\rm TDB}$ for this study. For many systems, there are literature data mixed in with the citizen science observations -- one of the initial goals of the ETD project was to have a single space for all amateur and professional transit observations \citep{poddany_exoplanet_2010}. The inclusion of literature results for any given planetary system in the database is nonetheless incomplete, with the majority posted in the earlier years of the project.\par

While the database is vast and thus brimming with potential, there are two reasons the ETD is limited in its immediate use: 1) the simplicity of its automatic transit fitting routine and 2) a submission process that allows for inconsistencies in the format of submitted data. 

The light curve fitting is done via a Levenberg-Marquardt (L-M) nonlinear least squares algorithm; namely,
\begin{equation}
\begin{split}
m(t_i) = A-2.5\log{F}(z[t_i,t_0,D,b],p,c_1) + \\ 
B(t_i-t_{mean})+C(t_i-t_{mean})^2,
\end{split}
\label{equation:eq1}
\end{equation}
where $m_i$ is the relative magnitudes taken at time $t_i$, and $F(z,p,c_1)$ is the relative flux, which depends on the radius ratio $p$, projected planet-star separation $z$, and limb-darkening coefficient $c_1$. Stellar limb-darkening is modeled linearly. The projected separation $z$ depends on the mid-transit time $t_0$, transit duration $D$, and impact parameter $b$. Systematic trends in the data are described by $B$ and $C$, and the zero-point shift of the magnitudes by $A$. In the fitting algorithm, $A$, $t_0$, $D$, $p$, $B$, and $C$ are free parameters. The linear limb-darkening coefficient $c_1$ is fixed at a value of 0.5 \citep{poddany_exoplanet_2010}.\par

Many studies using ETD observations re-fit the light curves \citep[e.g.][]{baluev_benchmarking_2015,mallonn_ephemeris_2019,davoudi_refined_2021,edwards_original_2021}. However, the different formats of the observations, both in terms of the machine-readable table format (comma-separated, tab-separated, etc.) and the timestamps provided, make reformatting and refitting the data a very time consuming process, particularly when assessing a large number of observations. This work involves a total of 4792 transit observations, making refitting the lightcurves an impracticable course of action in many instances.\par

Hence, instead of refitting the data, we take the fitted transit centres at face value, as the intent is to show that the ETD, and citizen science more generally, can be an efficient and convenient way to flag planetary systems with nonlinear long-term timing trends. 

The only data categories used in this study are the ETD times, the associated uncertainties on those times, and the epoch numbers. In doing this, there is an implicit assumption that the fitting routine used by ETD does not have a systematic bias and that the timestamps given are in the expected format of HJD$_{\rm UTC}$. If this is the case, we argue that the main effect of taking the data at face value is to increase the noise. This approach must be used with caution, as the ETD lightcurve fitting routine is not immune to systematic bias. The Levenberg-Marquardt algorithm is at higher risk of becoming trapped in a local minimum than a more rigorous MCMC routine, and there is no correction for red noise. Systematic noise in the lightcurves are modeled quadratically (Eq. \ref{equation:eq1}), though visual inspection of a subset of ETD transits implies that a higher-order approximation may improve the fitting. Additionally, limb-darkening is modeled linearly with a coefficient fixed at a value of 0.5 \citep{poddany_exoplanet_2010}, which \citet{petrucci_no_2013} found to be insufficient for the study of short term TTVs. Despite these potential sources of systematic bias, the agreement of our WASP-12 b analysis with the literature (Section \ref{sec:results}) shows that our assumption holds for the purpose of an efficient analysis of potential targets.\par

The overall quality of observations within the ETD is highly variable. Despite this, when analyzed carefully and consistently, they can be used in independent work and to supplement observations from professional observatories. A good example of the latter is in \citet{kipping_analysis_2011}, who directly compare TrES-2\,b transits from the ETD to high-precision Kepler observations of the same epoch number. The ETD observations were found to be in excellent agreement with the Kepler transits, with most of the transit centers matching to within $1-\sigma$.

To aid in analyzing the data, transit observations in the ETD are ranked by a ``data quality index'' (DQ) from 1-5, with 1 being those of the highest quality. The DQ index $\alpha$ is calculated as

\begin{equation}
\alpha = \frac{\delta}{S}\sqrt{\frac{N}{l}},
\label{equation:eq2}
\end{equation}
where $\delta$ is the minimum flux, $S$ the mean absolute deviation from the best-fit transit model, $N$ the number of data points, and $l$ the length of the observing run. The greater the value of $\alpha$,  the better quality data. \citet{poddany_exoplanet_2010} provides the thresholds used for ranking any given observation. For this study, only observations with data quality 1-3 were considered.\par

\section{Target Selection}\label{sec:targetselection}

There are over 400 systems in the ETD,\footnote{404 planets and 12,291 observations as of May 16, 2022} each with varying numbers of observations, data quality, and observational baselines. We are interested in data that might reveal the long-term evolution of planets and, given the range in data quality, focus only on systems that have a large number of observations.\par

The analysis is restricted to planetary systems that have more than 80 contributed observations with a data quality index of 3 or better. We found that model fits on systems with fewer observations become too dependent on local variations in the data. While there is not a clear transition, and a unique value will not exist, in this scenario a value of 80 removes systems with this strong dependency. Based on this criteria, 30 systems were chosen for further analysis, described further in Section \ref{sec:methods}. These 30 systems are listed in Table \ref{table:table1}.\par

%%% METHODS %%%

\section{Methods} \label{sec:methods}

A multi-step process was used to search for trends in the transit centre times. The transit centres are taken directly from the Exoplanet Transit Database (ETD) without performing new fits on the individual transit observations. The timestamps, provided by the database in HJD$_{\rm UTC}$, are converted to BJD$_{\rm TDB}$ for this study following the standard set by \citet{eastman_achieving_2010}. This conversion is done because, for example, the UTC (Universal Coordinated Time) standard drifts with the addition of leap seconds. Thus, comparing only UTC times that span multiple years may introduce an artificial drift in the transit times that could mimic a true variation in the HJ's orbit. The ETD includes long-term observations of WASP-12\,b, which serves as a critical reference due to its comprehensive study in the literature \citep{maciejewski_departure_2016,patra_apparently_2017,yee_orbit_2019,turner_decaying_2020,wong_tess_2022}. It is thus reassuring that the WASP-12\,b results from this study are in agreement with literature values, showing a highly significant decay (Table \ref{table:wasp12results}). This alone highlights the value of the ETD observations and suggests that some of the other compelling targets found in the ETD should be investigated further. Observations of select systems from the database are run through a Python pipeline developed specifically for this project. The data, results, and code used for this project are publicly available on Zenodo.\footnote{\url{https://doi.org/10.5281/zenodo.7098460}}\par

\begin{deluxetable}{lccccc}[t]
\tablecaption{Model comparison for initial analysis of 30 targets from the ETD}
\tablenum{1}
\label{table:table1}
\tablehead{\colhead{Target} & \colhead{Decay Rate} & \colhead{1-$\sigma$ Unc.} & \colhead{BIC$_{\rm linear}$} & \colhead{BIC$_{\rm decay}$} & \colhead{$\Delta$BIC} \\ 
\colhead{} & \colhead{(ms/yr)} & \colhead{(ms/yr)} & \colhead{} & \colhead{} & \colhead{} } 

\startdata
WASP-12\,b & -29.1 & 1.0 & 2693.4 & 2284.7 & -408.7 \\
TrES-2\,b & -20.7 & 2.1 & 1334.1 & 1287.3 & -46.8 \\
WASP-10\,b & -26.4 & 2.7 & 1890.9 & 1847.9 & -43.0 \\
HAT-P-32\,b & -30.2 & 3.3 & 1084.2 & 1044.5 & -39.7 \\
TrES-5\,b & -29.7 & 3.6 & 831.6 & 804.8 & -26.8 \\
TrES-3\,b & -5.08 & 0.63 & 2459.9 & 2433.2 & -26.7 \\
HAT-P-19\,b & -55.2 & 7.2 & 334.7 & 308.4 & -26.3 \\
TrES-1\,b & -7.9 & 1.4 & 656.5 & 646.6 & -9.9 \\
XO-2\,b & 13.0 & 3.1 & 490.5 & 487.1 & -3.4 \\
Qatar-1\,b & -6.1 & 1.4 & 1858.6 & 1855.3 & -3.3 \\
HAT-P-12\,b & -13.2 & 3.6 & 395.6 & 393.6 & -2.0 \\
GJ-436\,b & 11.6 & 3.3 & 466.5 & 465.1 & -1.4 \\
XO-1\,b & -13.3 & 3.9 & 265.6 & 264.5 & -1.1 \\
WASP-48\,b & -24.6 & 8.0 & 426.6 & 426.4 & -0.2 \\
WASP-52\,b & -11.8 & 3.7 & 513.6 & 513.4 & -0.2 \\
WASP-3\,b & -9.0 & 2.8 & 503.3 & 503.2 & -0.1 \\
HAT-P-10\,b & -15.3 & 1.8 & 636.0 & 636.9 & 0.9 \\
HAT-P-23\,b & -5.2 & 5.8 & 435.6 & 436.5 & 0.9 \\
HD189733\,b & 2.5 & 1.1 & 2452.3 & 2454.7 & 2.4 \\
WASP-2\,b & -4.2 & 1.9 & 641.7 & 644.3 & 2.6 \\
HAT-P-20\,b & -9.3 & 4.8 & 579.4 & 582.1 & 2.7 \\
WASP-4\,b & -1.09 & 0.66 & 356.4 & 359.1 & 2.7 \\
HAT-P-36\,b & 5.0 & 2.7 & 1099.5 & 1103.0 & 3.5 \\
CoRoT-2\,b & 2.2 & 1.7 & 649.9 & 653.5 & 3.6 \\
HAT-P-3\,b & -5.1 & 4.0 & 755.0 & 759.0 & 4.0 \\
GJ-1214\,b & -0.9 & 1.8 & 563.1 & 567.3 & 4.2 \\
HAT-P-37\,b & 6.6 & 6.8 & 431.6 & 435.9 & 4.2 \\
Qatar-2\,b & 1.6 & 2.2 & 681.4 & 685.6 & 4.2 \\
WASP-33\,b & -1.4 & 1.5 & 2926.9 & 2931.5 & 4.6 \\
WASP-43\,b & -0.7 & 1.2 & 744.7 & 749.5 & 4.8 \\
\enddata

\tablecomments{Results from the first run of the analysis pipeline on the raw transit centre data of 30 systems from the ETD. Targets are ranked by the likelihood of orbital decay model over a constant period model. The $\Delta$BIC values are such that a negative value favours the decay model.}
\end{deluxetable}

\subsubsection{Transit Timing Models}\label{sec:timingmodels}

Three different transit timing models are tested against the ETD transit centres using Markov Chain Monte Carlo (MCMC) methods. Initial tests were conducted using the \textit{emcee} package \citep{foreman-mackey_emcee_2013}. However, as discussed below, one of the models developed unstable behaviour and did not converge in general. To have more direct control over the MCMC algorithm, we implemented a custom Metropolis-Hastings MCMC with a Gibbs sampler \citep{ford_improving_2006}. This latter approach is used for all of the results presented here. For the models, we consider the cases of (1) a planet on a constant orbital period, (2) a planet on a decaying, circular orbit, and (3) a planet that has a constant orbital period, but the orbit is precessing.\par

In the constant orbital period case, the transit center times increase linearly with the transit epoch $E$ such that
\begin{eqnarray}
t_{tra} &=& t_0 + PE\rm~and\\ 
t_{occ} &=& t_0 + \frac{P}{2} + PE\rm,
\label{equation:eq3}
\end{eqnarray}
where $t_0$ is the reference epoch, $P$ is the period, and $t_{tra}$ and $ t_{occ}$ are the expected transit and occultation times, respectively.\par

The second model assumes a steady change in the orbital period with epoch. 
The simplest expression of this behaviour is to include a quadratic term in the expected transit centre times. If the fitted period derivative $\frac{dP}{dE}$ is negative, then the planet is decaying. The model is given by
\begin{eqnarray}
t_{tra} &=& t_0 + PE + \frac{1}{2}\frac{dP}{dE}E^2\rm~and\\
t_{occ} &=& t_0 + \frac{P}{2} + PE + \frac{1}{2}\frac{dP}{dE}E^2\rm.
\label{equation:eq4}
\end{eqnarray}

The third timing model attempts to use apsidal precession to explain timing variations, which requires the HJ orbit to have a non-zero eccentricity -- as we will see, even small eccentricities can give rise to measurable effects. To capture the system's behaviour in this model, the argument of pericentre $\omega$ is assumed vary at a constant rate, leading to sinusoidal trends in the timing data. Following \citet{patra_apparently_2017}, the transit and occultation times can be expressed as
\begin{eqnarray}
t_{tra} &=& t_0 + P_s E - \frac{e P_a}{\pi}\cos{\omega}\rm, \\
t_{occ} &=& t_0 + \frac{P_a}{2} + P_s E \frac{eP_a}{\pi}\cos{\omega}\rm,\\
\omega(E) &=& \omega_0 + \frac{d\omega}{dE}E\rm,~and \\
P_s &=& P_a\left(1-\frac{d\omega/{dE}}{2\pi}\right)\rm,
\label{equation:eq5}
\end{eqnarray}
for argument of pericentre $\omega$, phase $\omega_0$, and precession rate $d\omega/dE$. 
In these equations, $P_s$ represents the planet's sidereal period, which is assumed to be fixed, while $P_a$ is the ``anomalistic'' period. This latter period is also fixed, but accounts for the additional period signal due to precession.\par

The MCMC is used to determine the posterior distributions for the model parameters. The constant period model only has two free parameters: the reference epoch and the period. The decay model has three parameters, adding the $dP/dE$ term. Finally, the apsidal precession model has five parameters: the reference epoch, sidereal period, eccentricity, phase constant, and the precession rate.\par

Due to the different number of free parameters between the models, the Bayesian Information Criterion (BIC) is used to compare the relative accuracy and necessity of each model to describe the transit timing data. The BIC is defined as:
\begin{equation}
\text{BIC} = \chi^2 + k\log{n},
\label{equation:eq6}
\end{equation}
where $n$ is the number of data points and $k$ is the number of free parameters. Thus, the BIC accounts for the accuracy of the model through the $\chi^2$ statistic, while penalizing the model for the number of free parameters used to describe the data (i.e., the model's complexity). The difference in BIC relates to the Bayesian posterior odds ratio such that a $\Delta\text{BIC}=10$ corresponds to a 150:1 ratio for the stronger model, and a $\Delta\text{BIC}=5$ to 13:1 \citep{liddle_information_2007}.\footnote{It is important to note that the BIC is one of many valid model comparison statistics.} With this in mind, the evidence strongly supports the model with the lower BIC value over the one with the larger value when $\Delta\text{BIC}>10$. For $5<\Delta\text{BIC}<10$, the evidence supports the model with the lower BIC value, and if $\Delta\text{BIC}<5$, while the lower BIC model is favoured, the evidence in support of that distinction is weak.\par

\subsubsection{Data Cleaning}\label{sec:cleaning}

The transit observations in the ETD are of inconsistent quality and require at least some degree of cleaning. We note above that there is already a numerical data quality factor for each entry in the ETD, which can be used for making an initial selection of data.\footnote{As a reminder, we automatically remove transit centres that have an associated data quality factor of 4 and 5 (1 is the best and 5 is the worst)} However, this alone is insufficient due to spurious data points, i.e., points that are well-removed from emergent trends, but with a wide range of reported uncertainties. To address this, an iterative sigma-clipping algorithm is integrated into the MCMC analysis. The process is monitored to avoid introducing an artificial trend in the transit timing curves.\par

The pipeline begins by finding a preliminary best-fit orbital decay model using 100,000 iterations of the MCMC code. The first ten percent of the resulting chain is discarded as burn-in, which is found to be sufficient for the given chain lengths. The resulting preliminary model is subtracted from the timing data. The variance of the residuals is then calculated and used to flag any datum whose nominal value lies outside a 3-$\sigma$ deviation from the residual mean. Another round of fitting is then run, including the MCMC fitting, but with the flagged data excluded. This process is repeated until no points fall outside of the 3-$\sigma$ range. Convergence is achieved on average after $3.6$ iterations with an average number of $9$ observations removed. After this is complete a final, longer run of the MCMC is performed for the orbital decay model (5,000,000 iterations with 10\% discarded as burn-in) and, if desired, the precession model. \par

A worry of using the sigma-clipping method is that it may introduce a false signal, a potential issue that was recognized early on. To address this, during development of the fitting routines, the same overall method was tried but with a best-fit linear model subtracted from the data instead of the best-fit orbital decay model. It was found that the fitted parameters from both methods agreed within respective uncertainties and that the most likely timing model for any given system remained the same. Essentially, data that are better described by a linear trend just resulted in a quadratic best-fit model with very low curvature, and we did not ``clip'' the data into a quadratic trend. Given comparable results, we decided to focus on using the decay model for the sigma-clipping because it would be, in principle, more sensitive to trend detection. However, the linear model subtraction approach is still run to confirm strong timing trends and discussed as appropriate.\par

\begin{deluxetable*}{lcccc}
\tablecolumns{5}
\tablewidth{0pt}
\tablecaption{MCMC Priors}
\label{table:priors}
\tablehead{\colhead{Parameter} & \colhead{Symbol} & \colhead{Unit} & \colhead{Prior} & \colhead{Bounds}}
\startdata
\sidehead{\textbf{Constant Period Model}}
Transit Center & $t_{0}$ & BJD$_{\rm TDB}$ & Uniform & $(t_{ref}-0.5,t_{ref}+0.5)$ \vspace{0.05cm} \\ 
Period & $P$ & days & Uniform & $(P_{ref}-0.5,P_{ref}+0.5)$ \vspace{0.01cm} \\
\sidehead{\textbf{Orbital Decay Model}}
Transit Center & $t_{0}$ & BJD$_{\rm TDB}$ & Uniform & $(t_{best}-0.5,t_{best}+0.5)$ \vspace{0.05cm} \\
Period & $P$ & days & Uniform & $(P_{best}-0.5,P_{best}+0.5)$ \vspace{0.05cm} \\
Decay Rate & $dP/dE$ & days/epoch & Uniform & $(\num{-1e-7},\num{1e-7})$ \vspace{0.01cm} \\
\sidehead{\textbf{Apsidal Precession Model}}
Transit Center & $t_{0}$ & BJD$_{\rm TDB}$ & Uniform & $(t_{best}-a,t_{best}+a),a\in\{.01,.1\}$ \vspace{0.05cm} \\
Sidereal Period & $P_s$ & days & Normal & $(P_{best}-0.1,P_{best}+0.1)$ \vspace{0.05cm} \\
Argument of Periastron & $\omega_{0}$ & rad & Uniform & $(0,2\pi)$ \vspace{0.05cm} \\ 
Precession Rate & $d\omega/dE$ & rad/epoch & Log-Uniform & $(\num{1e-6},\num{1e-3})$  \vspace{0.05cm} \\
Eccentricity & $e$ & & Log-Uniform & $(\num{1e-5},\num{1e-1})$ \vspace{0.1cm} \\
\enddata
\vspace{0.02cm}\tablecomments{The parameters $t_{ref}$ and $P_{ref}$ are the reference ephemeris from the literature, whereas $t_{best}$ and $P_{best}$ are the best-fit values from sampling the constant period model. For the constant period and orbital decay models the bounds on $t_0$ and $P$ are generous to allow for exploration of a large parameter space, as there were no issues with converging on multiple solutions. The restrictions on the apsidal precession priors are discussed in section \ref{sec:modelfitting}}
\end{deluxetable*}

\subsubsection{Model Fitting Approach}\label{sec:modelfitting}
The data analysis pipeline described above is applied twice in total. For the first application, each of the 30 star-planet systems selected from the ETD (see Section \ref{sec:targetselection}) are processed in the pipeline. During this run,  all transit centres that have a DQ of 1, 2, or 3 are included in model fitting. Eight systems show evidence of a nonlinear trend (Table \ref{table:table1}) and are investigated further. Two additional targets are included due to potential nonlinear trends being previously reported in the literature (and with different studies reporting inconsistent results): WASP-4\,b \citep[see][]{bouma_wasp-4b_2019,southworth_transit_2019,baluev_homogeneously_2019,baluev_wasp-4_2020,bouma_wasp-4_2020,ivshina_tess_2022,turner_characterizing_2022} and WASP-43\,b \citep[see][]{blecic_spitzer_2014,murgas_gtc_2014,chen_broad-band_2014,ricci_multi-filter_2014,jiang_possible_2016,hoyer_ruling_2016,zhao_photometric_2018}.

The resulting ten systems are then manually cleaned before being run through the data pipeline a second time. This is a time-consuming process, which is why it is only done at this stage. Specifically, any observation in the database without a clear transit or a clear ingress and egress are manually flagged and excluded from further analysis. Entries without a reported uncertainty are also excluded, automatically. Because this project purposefully does not involve re-fitting light curves, the manual data inspection is necessary to filter out transits without clearly defined edges. It also demonstrates that relying on the DQs alone is insufficient, despite the intended use of DQs. For example, \citet{poddany_exoplanet_2010}  state that partial transits uploaded to the ETD would automatically be given a data quality index of 5, but this automated flagging appears to have stopped after some time. The transit observations that are included in these cleaned datasets are listed in Appendix \ref{sec:appendixB} for reproducibility. With these ten systems now having been manually cleaned, the data are run through the pipeline a second time. At this point, the apsidal precession model is also included.\par

The orbital decay model fitting is well-behaved and consistently converges to a clear solution. Uniform prior distributions are placed on the 3 free parameters ($t_0$, $P$, and ${dP/{dE}}$) and a wide parameter space can be explored. Fitting the precession model, however, proved to be non-trivial due to degeneracies between viable models, which we attribute to be due, in part, to the high variance of the data. In particular, the anti-correlations between the eccentricity, $e$, and the precession rate, ${d\omega/{dE}}$, as well as between $e$ and the reference transit time, $t_0$, lead to posterior solutions that show two or more strong peaks. After exploring many options, a selection of priors and bounds on parameters was determined to be necessary to derive converged results. Indeed, it is this issue that ultimately led to writing a custom MCMC routine. To ensure convergence of our posterior chains, we examined a number of outputs such as trace plots, corner plots, and the autocorrelation coefficient as a function of lag, all of which are consistent with convergence.

Table \ref{table:priors} summarizes the prior distributions and bounds placed on the parameters for all three transit timing models. A normal distribution prior is placed on the period, centered on the best-fit result from the constant period model fit. This is justified by the observation that this constant period can be viewed as an analog for the sidereal period in the apsidal precession model (Equation \ref{equation:eq5}). A uniform prior distribution is appropriate for $\omega_0$, and log-uniform priors are used for $e$ and ${d\omega/{dE}}$ to account for the wide range of possible parameter space.\par 

Without placing a formal bound on the precession rate, the model tends to go to high frequencies in an attempt to fit the datum-to-datum variation, an unphysical situation. To address this, we use the results of \citet{ragozzine_probing_2009}, who estimate that WASP-12\,b should have a precession rate of 19 deg/yr (roughly 0.001 radian/epoch). Out of all of the systems they studied, WASP-12\,b has by far the highest predicted rate. With this in mind, an upper bound of 0.001 radian/epoch is placed on ${d\omega/{dE}}$ for all systems, even for WASP-12\,b, as  a higher bound on the precession rate does not result in a better fit to the data in this case (tests using a higher bound for WASP-12\,b were conducted to confirm this result). A lower bound of \num{1e-6} rad/epoch is also chosen to avoid the M-H algorithm becoming trapped in a local minimum, as very low values of ${d\omega/{dE}}$ and $e$ cause the fitted precession model to approach a line even when the results of the orbital decay model indicate the presence of curvature in the data. The eccentricity is restricted to be less than 0.1, which is appropriate for the HJ population, and greater than \num{1e-5}. \citet{ragozzine_probing_2009} found that even eccentricities on the order of $10^{-5}$ can result in detectable transit timing variations, though measurements of eccentricities are essentially unconstrained by $10^{-3}$.\par

The strictest requirement for convergence of the apsidal precession model is on the parameter bounds of the reference transit center time $t_0$. The parameter $t_0$ is allowed to be within a range of $\pm0.01$ (just over 30 minutes) or $\pm0.1$ days around the best-fit value from the constant period model, depending on the system. This final adjustment allows convergence on the solution, but only as a perturbation of the constant period case. To aid convergence, a non-linear least squares fit (\textit{scipy} implementation) of the precession model was done within these bounds to select the best initial value of the model parameters. Altogether, this procedure facilitates convergence of the precession models with the given data sets, and with all of this in mind, we next discuss the results.\par

%%% RESULTS %%%

\section{Results} \label{sec:results}
As noted above, the first run of the analysis pipeline includes all DQ 1-3 observations from all 30 of the initial targets (see Section \ref{sec:targetselection}). Table \ref{table:table1} ranks the star-planet systems by the difference in their Bayesian Information Criterion ($\Delta\text{BIC}$), with a negative value favouring the decay model. The best-fit period decay rates, along with the corresponding 1-$\sigma$ uncertainties, are given in $\rm ms~yr^{-1}$. For some systems, such as GJ 436\,b and XO-2\,b, the period derivative is positive. In such cases, we found the scatter in the data to be very high, and did not investigate such systems further in this study. They will nonetheless need to be examined in future studies as better data are acquired.\par

From Table \ref{table:table1}, the top eight candidates from this first analysis are selected for the second, deeper analysis described in Section \ref{sec:modelfitting}. Those top eight candidates, in order of evidence for decay, are WASP-12\,b, TrES-2\,b, WASP-10\,b, HAT-P-32\,b, TrES-5\,b, TrES-3\,b, HAT-P-19\,b, and TrES-1\,b. With the exception of TrES-1\,b, all of these systems yield a $\Delta\text{BIC}>10$ (Section \ref{sec:timingmodels}). The evidence for TrES-1\,b is still compelling, with $\Delta\text{BIC}=9.9$. In addition to these eight targets, WASP-4\,b and WASP-43\,b are selected due to their relevance in the literature, for a total of ten targets.\par

The results of the second round of analysis is summarized in Table \ref{table:table2}. In the following sub-sections, each of these star-planet systems is presented individually, with an expanded discussion of that system's model fitting. The cleaned data sets for the top ten systems are further provided in the tables listed in Appendix \ref{sec:appendixB}, including the epoch number, transit centres and respective uncertainty, data quality index, and observer name(s). The full MCMC outputs from the analyses can be found in the tables in Appendix \ref{sec:appendixA}, which include the best-fit model parameters and their 1-$\sigma$ uncertainties, as well as metrics such as the $\chi^2$, number of data, etc.\par

Out of the eight highest-ranked systems from Table \ref{table:table1}, all but TrES-3\,b maintain statistically significant evidence of orbital decay in the ETD transit centres. WASP-4\,b and WASP-43\,b, both of which favoured a constant period transit timing model in the first analysis (Table \ref{table:table1}), exhibit varying results. For WASP-43\,b, the second analysis maintains that the planet's transits follow a linear ephemeris (see Section \ref{sec:wasp43}), whereas for WASP-4\,b the favoured model changes to orbital decay with a decay rate of $-6.21 \pm 0.70$ ms/yr (see Section \ref{sec:wasp4}).\par

As a cautionary approach, the secondary analysis was repeated under the assumption that the transit centre uncertainties in the ETD are underestimated, thus replacing them with the standard deviation of the spread of the timing residuals -- i.e., the observed transit centres minus the times calculated from the best-fit linear ephemeris. In general, using a nonlinear least squares algorithm can lead to optimistic parameter uncertainties. In addition, when describing the ETD lightcurve fitting routine, \citet{poddany_exoplanet_2010} acknowledge that the errors of their L-M fit may be underestimated due to the lack of red noise correction and a fixed impact parameter. Substituting the ETD uncertainties in this way allow for an examination of the curvature in the data unbiased by the uncertainties given by the ETD lightcurve fitting tool (Table \ref{table:table3}).

It is notable that, with the exception of HAT-P-32\,b and WASP-10\,b, all of the resulting decay rates agree within the respective uncertainties with those in Table \ref{table:table2}. However, as to be expected, the uncertainties are much larger, and only three systems (WASP-12\,b, HAT-P-19\,b, and TrES-1\,b) still exhibit statistical evidence for departure from a linear ephemeris, using a fixed variance of the data (\ref{table:table3}).\par

\begin{deluxetable*}{lcccccc}
\tablenum{2}
\label{table:table2}
\tablecaption{Model comparison for secondary analysis of top 10 targets}
\tablehead{\colhead{Target} & \colhead{Decay Rate} & \colhead{1-$\sigma$ Unc.} & \colhead{BIC$_{\rm linear}$} & \colhead{BIC$_{\rm decay}$}& \colhead{$\Delta$BIC} & \colhead{BIC$_{\rm precession}$} \\ 
\colhead{} & \colhead{(ms/yr)} & \colhead{(ms/yr)} & \colhead{} & \colhead{} & \colhead{} } 

\startdata
WASP-12\,b & -31.6 & 1.0 & 2353.1 & 1911.8 & -441.3 & 1929.4 \\
WASP-4\,b & -6.21 & 0.70 & 404.9 & 369.5 & -35.4 & 378.8 \\
WASP-10\,b & -21.9 & 2.4 & 1011.3 & 976.0 & -35.3 & 984.6 \\
HAT-P-19\,b & -57.7 & 7.3 & 270.4 & 243.1 & -27.3 & 251.5 \\
TrES-5\,b & -34.5 & 4.6 & 541.1 & 518.9 & -22.2 & 529.0 \\
TrES-1\,b & -10.9 & 2.0 & 190.9 & 181.2 & -9.7 & 190.2 \\
TrES-2\,b & -12.6 & 2.4 & 878.9 & 870.6 & -8.3 & 882.4 \\
HAT-P-32\,b & -7.3 & 1.5 & 677.2 & 669.8 & -7.4 & 679.2 \\
TrES-3\,b & -2.75 & 0.78 & 939.8 & 938.9 & -0.9 & 982.7 \\
WASP-43\,b & -1.0 & 1.2 & 576.2 & 580.6 & 4.4 & 593.1 \\
\enddata

\tablecomments{Transit timing model comparison for the second run of the analysis pipeline on the reduced data of the top 10 systems of interest. The $\Delta$BIC is calculated for the comparison of the constant-period and orbital decay transit timing models, where a negative value favours the latter. The BIC values for the apsidal precession model fits are provided for further comparison.}
\end{deluxetable*}

\begin{deluxetable*}{lccccc}
\tablecaption{Model comparison for secondary analysis - data variance for ETD transit times}
\tablenum{3}
\label{table:table3}
\tablehead{\colhead{Target} & \colhead{Decay Rate} & \colhead{1-$\sigma$ Unc.} & \colhead{BIC$_{\rm linear}$} & \colhead{BIC$_{\rm decay}$} & \colhead{$\Delta$BIC} \\ 
\colhead{} & \colhead{(ms/yr)} & \colhead{(ms/yr)} & \colhead{} & \colhead{} & \colhead{} } 

\startdata
WASP-12\,b & -34.8 & 4.9 & 223.7 & 202.7 & -21.0 \\
HAT-P-19\,b & -64 & 17 & 80.6 & 78.2 & -2.4 \\
TrES-1\,b & -16.0 & 3.7 & 73.3 & 68.3 & -5.0 \\
WASP-4\,b & -6.7 & 2.4 & 62.6 & 62.7 & 0.1 \\
TrES-2\,b & -22.0 & 8.0 & 159.0 & 160.3 & 1.3 \\
TrES-5\,b & -25 & 11 & 118.3 & 120.5 & 2.2 \\
HAT-P-32\,b & -32 & 12 & 95.5 & 96.7 & 1.2 \\
WASP-10\,b & -10.1 & 7.6 & 131.6 & 135.5 & 3.9 \\
WASP-43\,b & 3.5 & 4.0 & 126.5 & 130.9 & 4.4 \\
TrES-3\,b & 0.01 & 1.9 & 227.8 & 233.1 & 5.3 \\
\enddata

\tablecomments{Model comparison of the results from the reduced data of the top 10 systems after replacing the ETD transit centre uncertainties with the standard deviation of the nominal timing residuals. A negative $\Delta$BIC value favours the orbital decay model.}
\end{deluxetable*}

\subsubsection{WASP-12\,b}\label{sec:wasp12}
\begin{figure*}
\includegraphics[width=0.9\textwidth]{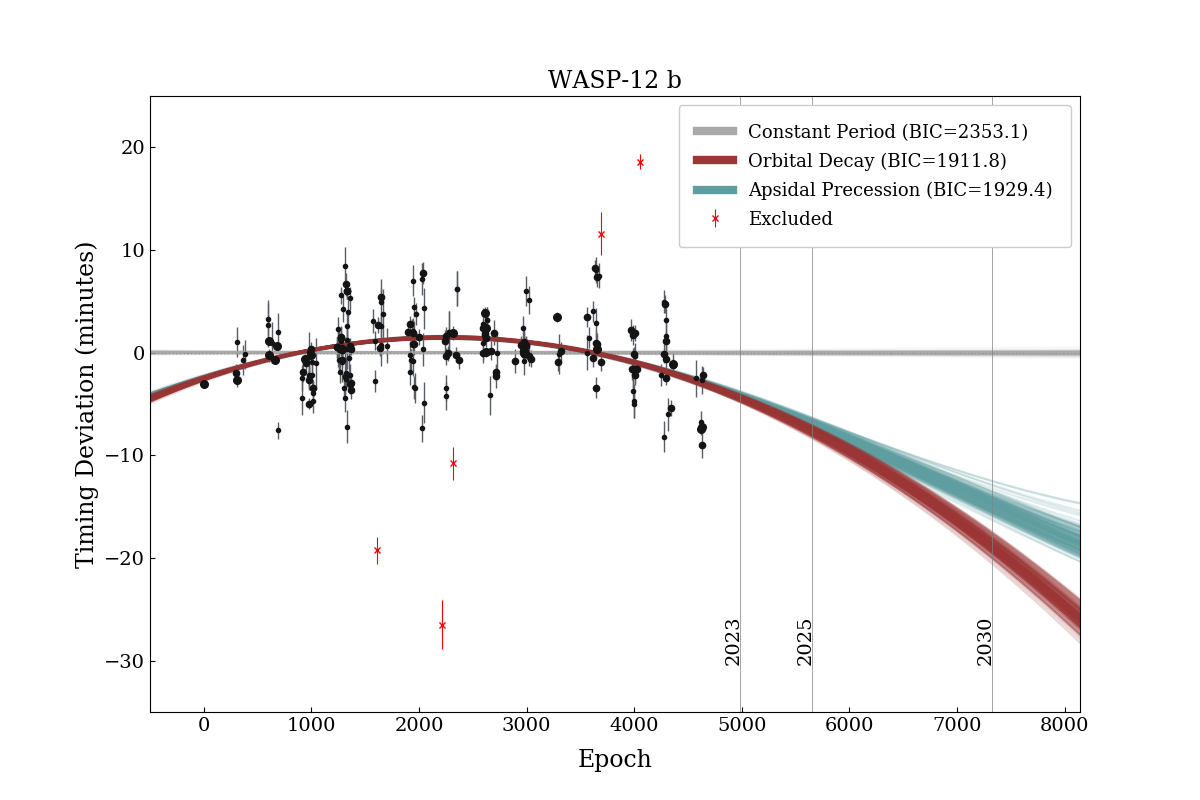}
\includegraphics[width=0.9\textwidth]{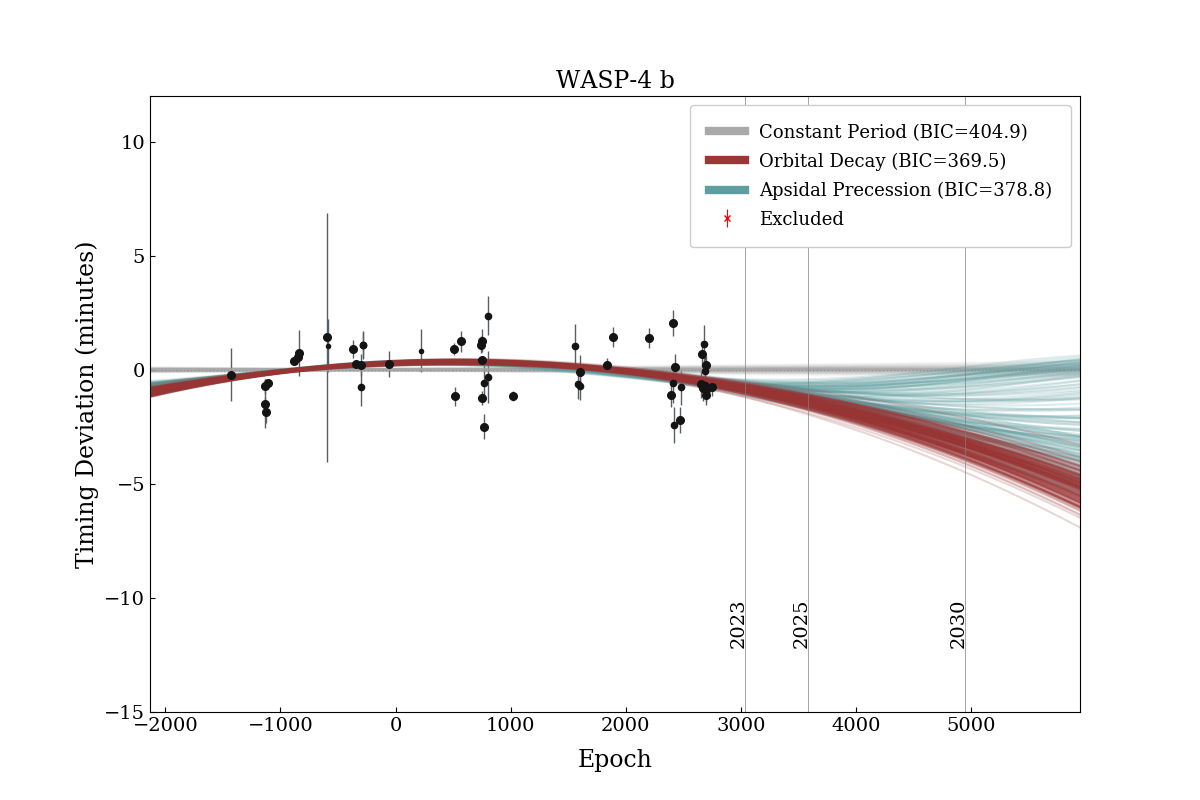}
\caption{Timing residuals of WASP-12\,b (top) and WASP-4\,b (bottom) with future projections of the three transit timing models shown with 150 random draws from the MCMC posterior chains. Note that for aesthetic reasons, the scaling of the y-axis varies for each plot. Each datum is the difference between an observed time and the time predicted by the best-fit constant period model. The size of the data points correspond to the data quality index from 1-3, with the higher-quality transits being larger. The red crosses represent observations removed during the sigma-clipping process.}
\label{fig:wasp12wasp4}
\end{figure*}

WASP-12\,b is a $1.4$ M\textsubscript{J}, $1.8$ R\textsubscript{J} Hot Jupiter discovered in 2008 on a $1.09142090 \pm 0.0000002$ day orbit \citep{hebb_wasp-12b_2008,bonomo_gaps_2017}. The transit timing variations of WASP-12\,b have been extensively studied, as it is the only exoplanet for which orbital decay has been unambiguously detected. The first suggestion of a period drift was from \citet{maciejewski_departure_2016}, who found that the WASP-12\,b transit times diverged from a linear ephemeris. Various studies have since confirmed this finding, all with compatible decay rates \citep[see e.g.][]{patra_apparently_2017, yee_orbit_2019,turner_decaying_2020}, the most recent result being $-30.27 \pm 1.11$ ms/yr from \citet{ivshina_tess_2022}.\par 

At the time of writing, the ETD hosts 295 observations of WASP-12\,b spanning the years 2008 to 2021, 257 of which have DQ 1-3. The curvature of the transit timing residuals is clear when looking at the ETD generated O-C plot.\footnote{\url{http://var2.astro.cz/ETD/etd.php?STARNAME=WASP-12&PLANET=b}} Prior to any data cleaning, the ETD transit times fit an orbital decay model with a period derivative of $-29.1 \pm 1.0$ ms/yr, preferred over a linear ephemeris by $\Delta\text{BIC}$ of $408.7$ (Table \ref{table:table1}). This is by far the greatest $\Delta\text{BIC}$ of all of the targets, with the next best being $46.8$ for TrES-2\,b.\par 

After removing partial and duplicate transits, 218 WASP-12\,b transit times remain (Table \ref{table:wasp12data}), five of which are from published literature studies. The final analysis of the cleaned WASP-12\,b data increases the $\Delta\text{BIC}$ to $441.3$ favouring orbital decay at a rate of $-31.6 \pm 1.0$ ms/yr (see Figure \ref{fig:wasp12wasp4}). The quadratic trend is also favoured over apsidal precession by a $\Delta\text{BIC}$ of $17.6$. The best-fit precession rate is $0.000336^{+0.000026}_{-0.000013}$ rad per epoch with an eccentricity of $0.0287^{+0.0022}_{-0.0039}$ (Table \ref{table:wasp12results}).\par 

\subsubsection{WASP-4\,b}\label{sec:wasp4}

The Hot Jupiter WASP-4\,b has an orbital period of $1.338231624 \pm 0.000000068$ days and a radius and mass of $1.42$ R\textsubscript{J} and $1.22$ M\textsubscript{J}, respectively \citep{wilson_wasp-4b_2008,bonomo_gaps_2017}. The first suggestion of a period variation in WASP-4\,b in the literature was from \citet{bouma_wasp-4b_2019}, who added 18 TESS transits to a collection of observations going back to 2007 and found the period to be changing at a rate of $-12.6 \pm 1.2$ ms/yr. Later in the same year, \citet{southworth_transit_2019} included their own data from the years 2009-2019 and found that WASP-4\,b is decaying at a rate of $-9.2 \pm 1.1$ ms/yr, while \citet{baluev_homogeneously_2019} conducted a homogeneous analysis and found no statistically significant evidence of a nonlinear timing trend. In 2020, \citet{baluev_wasp-4_2020} detected a nonlinear trend with 3.4-$\sigma$ significance and \citet{bouma_wasp-4_2020} reported a period change of $-8.64 \pm 1.26$ ms/yr for WASP-4\,b from transit observations. However, after an analysis of new radial velocity data, they found that the Doppler effect from WASP-4\,b accelerating toward the earth has the effect of decreasing the period by $-5.94 \pm 0.39$ ms/yr. Thus, they conclude the period change is `mostly or entirely' due to acceleration of the system along the line of sight \citep{bouma_wasp-4_2020}. However, a recent comprehensive analysis from \citet{turner_characterizing_2022} did not detect this acceleration in the RV data, instead finding evidence of an additional planet in the system with a period of approximately 7000 days in addition to TTVs that are consistent with orbital decay at a rate of $-7.33 \pm 0.71$ ms/yr. Another recent study from \citet{ivshina_tess_2022} measured the period of WASP-4\,b to be changing at a compatible rate of $-5.81 \pm 1.58$ ms/yr.

Like WASP-43\,b, WASP-4\,b was selected for further investigation in this study because of this interest in the literature. The ETD lists 68 transit observations of WASP-4\,b from the years 2009-2021, 63 of which are DQ 1-3. As many of the transits are high quality, this is a valuable resource for contributing to our understanding of the WASP-4 star-planet system. The initial analysis of the ETD transit centres does not support a departure from a linear ephemeris, with the $\Delta\text{BIC}$ favouring the constant period model by $2.7$ and the best-fit decay rate being $-1.09 \pm 0.66$ ms/yr (Table \ref{table:table1}). After the exclusion of partial and duplicate transits for the second analysis, 55 observations of WASP-4\,b remain (Table \ref{table:wasp4data}), ten of which are from published literature studies.

The analysis of the cleaned WASP-4\,b data changes the favoured transit timing model (Figure \ref{fig:wasp12wasp4}). In the second analysis, orbital decay at a rate $-6.21 \pm 0.70$ ms/yr is strongly favoured for WASP-4\,b over the constant period model by a $\Delta\text{BIC}$ of $35.4$ (Table \ref{table:table2}). The apsidal precession model is also a clear contender by a $\Delta\text{BIC}$ of $26.1$ when compared to a constant period, but orbital decay is still strongly favoured over precession by $\Delta\text{BIC}=9.3$. The best-fit precession rate is $0.00048^{+0.00040}_{-0.00024}$ with an eccentricity of $0.0027^{+0.0079}_{-0.0017}$ (Table \ref{table:wasp4results}). This study does not evaluate the likelihood or directly model a line-of-sight acceleration of the WASP-4\,b system, but in this context, the orbital decay rate can be treated as a constant period derivative within a different physical context. Whatever the underlying reason, the evidence in the data support a timing variation. 

\subsubsection{WASP-10\,b}\label{sec:wasp10}

\begin{figure*}
\includegraphics[width=0.9\textwidth]{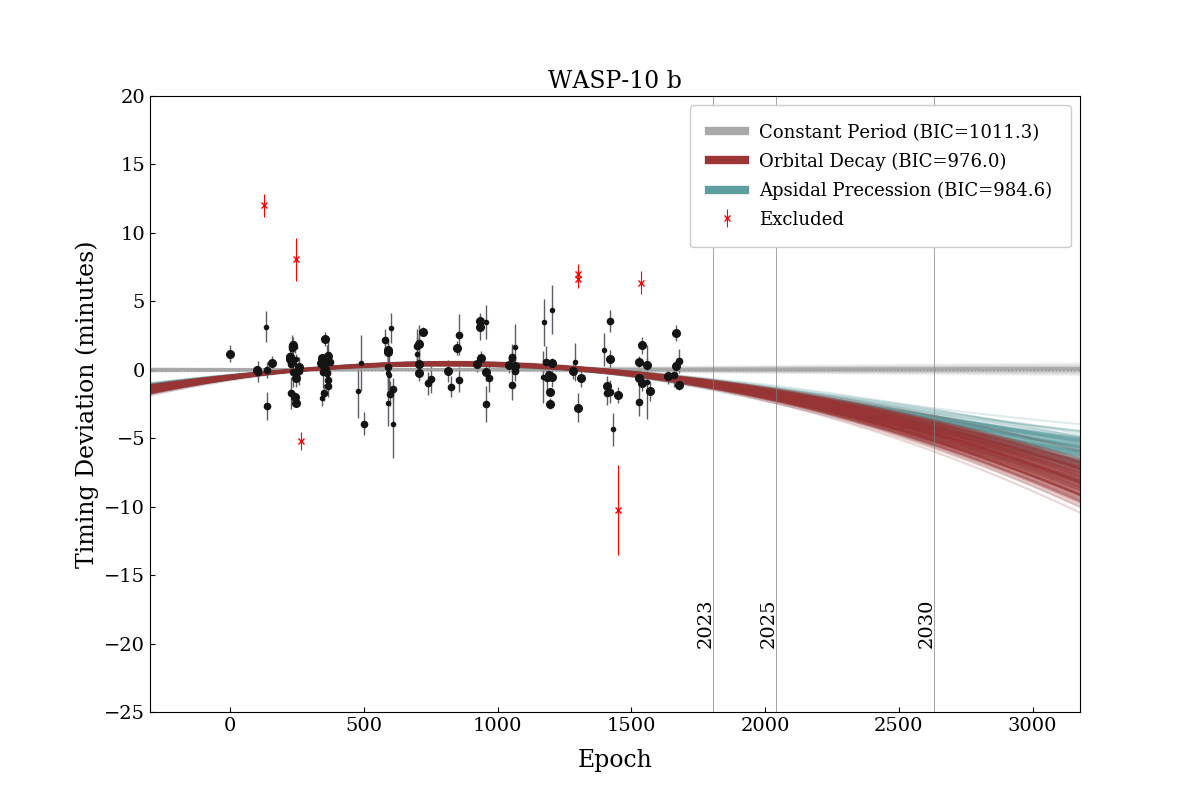}
\includegraphics[width=0.9\textwidth]{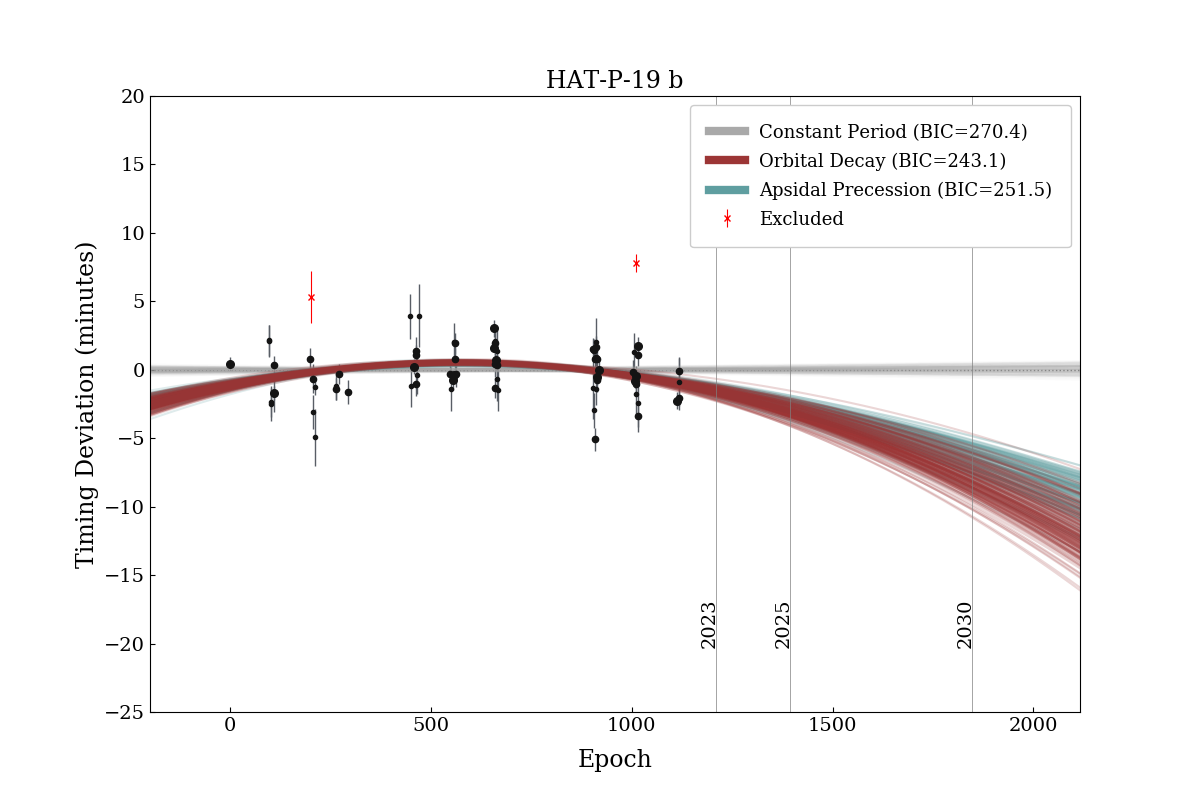}
\caption{Similar to Figure \ref{fig:wasp12wasp4}, but for WASP-10\,b (top) and HAT-P-19\,b (bottom).}
\label{fig:wasp10hatp19}
\end{figure*}

WASP-10\,b is a $3.2$ M\textsubscript{J}, $1.1$ R\textsubscript{J} Hot Jupiter discovered on a $3.09272932 \pm 0.00000032$ day orbit in 2009 \citep{christian_wasp-10b_2009,bonomo_gaps_2017}. It has been the subject of multiple timing studies in the past due to a measured periodic TTV, with proposed explanations including starspot occultations \citep{barros_transit_2013} and a $0.1$ M\textsubscript{J} companion with a period of about $5.23$ days \citep{maciejewski_transit_2011,maciejewski_analysis_2011}. A second planet in the system has never been confirmed, and we have not found further discussion in the literature of other studies looking for long-term trends in the transit times of WASP-10\,b, such as orbital decay or apsidal precession.\par

The ETD hosts 202 observations of WASP-10\,b spanning the years 2007-2021, 174 of which are DQ 1-3. An initial examination of the WASP-10\,b data gave a best-fit orbital decay rate near $-130$ ms/yr, largely driven by a single data point from \citet{johnson_smaller_2009}. Removing this observation from the data set gave a more conservative period derivative of $-26.4 \pm 2.7$ ms/yr, which is the value reported in the initial round of analyses (Table \ref{table:table1}). WASP-10\,b was selected for further analysis, with the orbital decay model being strongly favoured over a constant period by a $\Delta\text{BIC}$ of $43.0$.\par

During the data reduction process it was revealed that the observation in \citet{johnson_smaller_2009} was corrected in an erratum \citep{johnson_erratum_2010}. In addition to this, the transit centres published in \citet{christian_wasp-10b_2009} and \citet{maciejewski_transit_2011} had to be converted to BJD$_{\rm TDB}$ from BJD$_{\rm UTC}$ and BJD$_{\rm TT}$, respectively. After removing all partial and duplicate transits, there are 129 observations of WASP-10\,b remaining, including 19 measurements from the literature (Table \ref{table:wasp10data}). The final analysis of the cleaned WASP-10\,b data retained the strong preference for the orbital decay model (Figure \ref{fig:wasp10hatp19}) by a $\Delta\text{BIC}$ of $35.3$ with a best-fit decay rate of $-21.9 \pm 2.4$ ms/yr (Table \ref{table:table2}). The apsidal precession model is also a better fit to the data than a constant period, but orbital decay is favoured by $\Delta\text{BIC}$ of $8.6$. The best-fit apsidal precession model gives an eccentricity $0.0037^{+0.0027}_{-0.0011}$ and precession rate $0.00078^{+0.00015}_{-0.00020}$ rad per epoch (Table \ref{table:wasp10results}).\par 

\subsubsection{HAT-P-19\,b}\label{sec:hatp19}

HAT-P-19\,b is a Saturn-mass ($0.29$ M\textsubscript{J}) planet with a $1.1$ R\textsubscript{J} radius on a $4.008778 \pm 0.000006$ day orbit \citep{hartman_hat-p-18b_2011,bonomo_gaps_2017}. In the discovery paper, \citet{hartman_hat-p-18b_2011} detected a linear trend in the radial velocity residuals, indicating the presence of a another body in the planetary system. This observation led to multiple studies seeking periodic TTVs \citep{seeliger_ground-based_2015,maciejewski_new_2018,basturk_holistic_2020}, but we have not found further discussion in the literature on testing for long-term trends in the timing residuals.\par 

HAT-P-19\,b was selected as a target of interest after the first round of analyses showed a preference for orbital decay ($\Delta\text{BIC}=26.3$) at a rate of $-55.2 \pm 7.2$ ms/yr (Table \ref{table:table1}). The ETD has 98 observations of HAT-P-19\,b spanning the years 2009-2021, 88 of which are DQ 1-3. The only literature observation is the discovery epoch, which had to be converted from BJD$_{\rm UTC}$ to BJD$_{\rm TDB}$ \citep{hartman_hat-p-18b_2011}. After the exclusion of partial and duplicate transits, 75 observations of HAT-P-19\,b remain (Table \ref{table:hatp19data}). The second analysis of HAT-P-19\,b on this cleaned data set (Figure \ref{fig:wasp10hatp19}) yields a period derivative of $-57.7\pm 7.3$ ms/yr, this time favouring the decay model by a $\Delta\text{BIC}$ of $27.3$ (Table \ref{table:table2}). The orbital decay model is also favoured over apsidal precession by a $\Delta\text{BIC}$ of $8.4$. The best-fit precession model yields an eccentricity of $0.0068^{+0.0009}_{-0.0010}$ and precession rate $0.000905^{+0.000059}_{-0.000068}$ rad per epoch (Table \ref{table:hatp19results}).\par 

\subsubsection{TrES-5\,b}\label{sec:tres5}
\begin{figure*}
\includegraphics[width=0.9\textwidth]{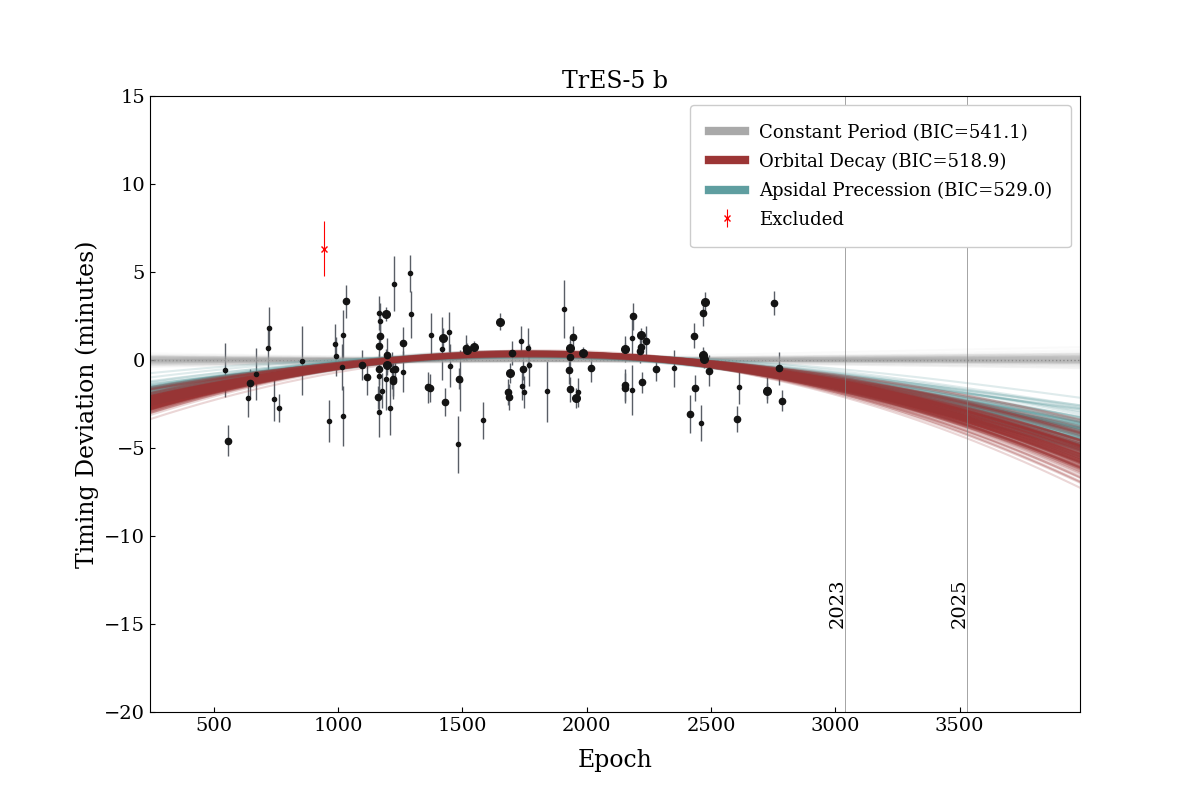}
\includegraphics[width=0.9\textwidth]{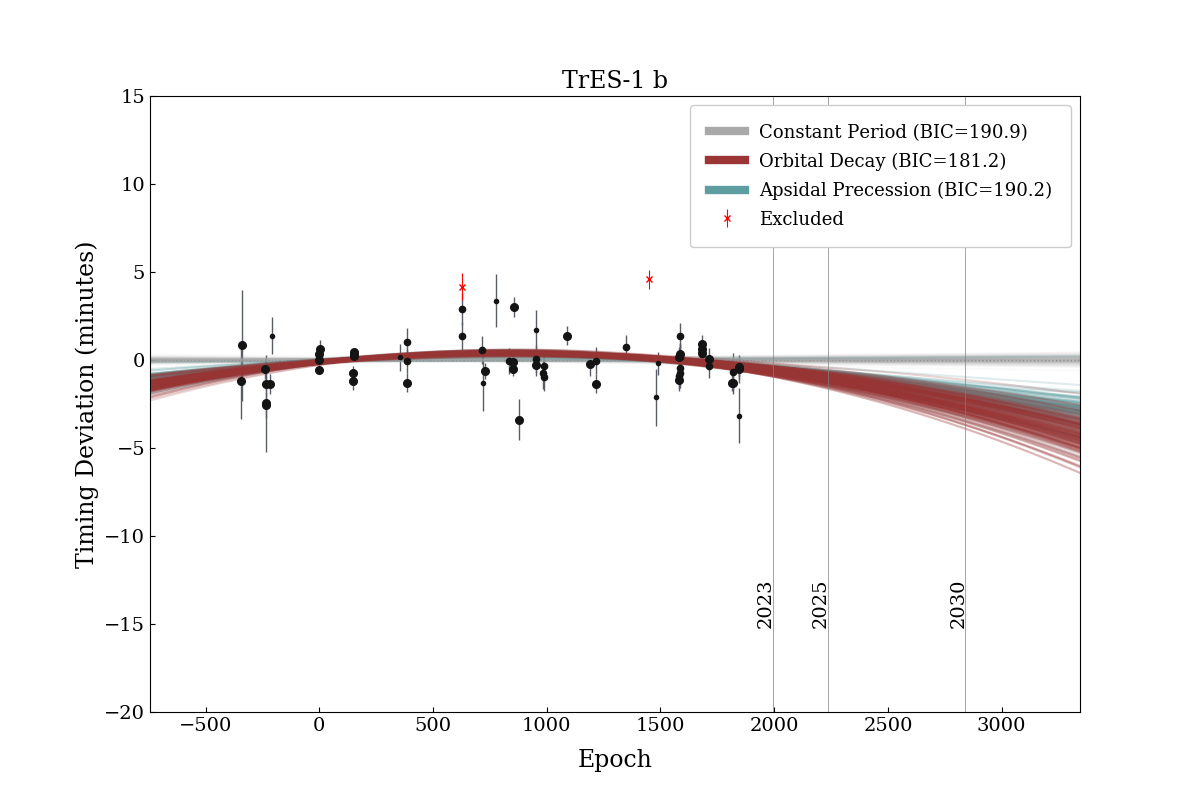}
\caption{Similar to Figure \ref{fig:wasp12wasp4}, but for TrES-5\,b (top) and TrES-1\,b (bottom).}
\label{fig:tres5tres1}
\end{figure*}

TrES-5\,b is a $1.8$ M\textsubscript{J}, $1.2$ R\textsubscript{J} Hot Jupiter discovered in 2011 on a $1.48224460 \pm 0.00000070$ day orbit \citep{mandushev_tres-5_2011,bonomo_gaps_2017}. The first suggestion of transit timing variations in the literature was by \citet{sokov_transit_2018}, who detected periodic timing variations indicative of a second planet in the system with a mass of $0.24$ M\textsubscript{J} in the 1:2 resonance orbit. \citet{maciejewski_revisiting_2021} were not able to confirm or reject the existence of an additional body in the system via short-term TTVs. However, they found a long-term variation of the orbital period of $-20.4 \pm 4.7$ ms/yr and conclude it is most likely a line-of-sight acceleration of the system induced by a massive, wide-orbiting companion. At the time of writing, the most recent timing study of TrES-5\,b is from \citet{ivshina_tess_2022}. Their analysis, which includes TESS data, found the period to be changing at a rate $-17.47 \pm 3.79$ ms/yr.\par 

The first analysis of the TrES-5\,b ETD transit centres indicates the period could be changing at a rate of $-29.7 \pm 3.6$ ms/yr, with the quadratic orbital decay model favoured over a constant period by a $\Delta\text{BIC}$ of $26.8$ (Table \ref{table:table1}). The ETD hosts 268 transits of TrES-5\,b from the years 2012-2021, 233 of which are DQ 1-3. There are no literature observations of TrES-5\,b on the database. After removal of partial and duplicate transits there are 110 remaining (Table \ref{table:tres5data}). The analysis of these cleaned data showed an increase in the best-fit period derivative to $-34.5 \pm 4.6$ ms/yr in the orbital decay model, favoured over a constant period by a $\Delta\text{BIC}$ of $22.2$  (Table \ref{table:table2}). The quadratic model is also a better fit to the data than the apsidal precession model by a difference $\Delta\text{BIC}=10.1$ (see Figure \ref{fig:tres5tres1}). The best-fit precession rate is $0.00058^{+0.00021}_{-0.00014}$ rad per epoch with an eccentricity $0.0098^{+0.0068}_{-0.0045}$ (Table \ref{table:tres5results}). 

\subsubsection{TrES-1\,b}\label{sec:tres1}

TrES-1\,b is a $0.7$ M\textsubscript{J}, $1.1$ R\textsubscript{J} Hot Jupiter on a $3.03006973 \pm 0.00000018$ day orbit \citep{alonso_tres-1_2004,bonomo_gaps_2017}. TrES-1\,b was discovered in 2004, the first of the Trans-Atlantic Exoplanet Survey (TrES), and thus has one of the longest potential observational baselines in this study. It has been the subject of various short-term TTV studies in the past \citep{rabus_transit_2008,rabus_transit_2009,baluev_benchmarking_2015}, but no significant variations were detected. Most recently \citet{ivshina_tess_2022} have included TESS data and found the TrES-1\,b period to be changing at a rate $-18.36 \pm 3.73$ ms/yr. As this was part of a much larger study they did not go into further detail, but suggest the system is worth monitoring. 

The ETD hosts 223 observations of TrES-1\,b spanning the years 2004-2021, 169 of which are DQ 1-3. Prior to any data cleaning, the ETD transit times fit an orbital decay model with a period derivative of $-7.9 \pm 1.4$ ms/yr, preferred over a linear ephemeris by $\Delta\text{BIC}$ of $9.9$ (Table \ref{table:table1}). The TrES-1\,b system has the weakest evidence of a nonlinear timing trend out of the top eight systems picked from Table \ref{table:table1}, but the results still fit the criteria for strong evidence with a $\Delta\text{BIC}>5$. Nineteen of the TrES-1\,b transit times on the database are from published literature studies, though it was discovered that five of them are duplicates and were excluded. In addition, the observations from \citet{hrudkova_searching_2008} had to be converted from BJD$_{\rm UTC}$ to BJD$_{\rm TDB}$. After these considerations, there are 68 remaining observations (Table \ref{table:tres1data}). 

The final analysis of the cleaned TrES-1\,b data (see Figure \ref{fig:tres5tres1}) yields a best-fit period derivative of $-10.9 \pm 2.1$ ms/yr with the orbital decay model being favoured over a constant period by a $\Delta\text{BIC}$ of $9.7$ (Table \ref{table:table2}). Thus, the results of the second analysis are consistent with the first, which contained data duplicates and observations with inconsistent timestamps, suggesting the fit results are robust. The BIC values for the constant period and apsidal precession models are almost indistinguishable (Table \ref{table:tres1results}). The best-fit apsidal precession model gives an eccentricity $0.0030^{+0.0037}_{-0.0015}$ and precession rate $0.00057^{+0.00024}_{-0.00019}$ rad per epoch.\par 

\subsubsection{TrES-2\,b}\label{sec:tres2}

\begin{figure*}
\includegraphics[width=0.9\textwidth]{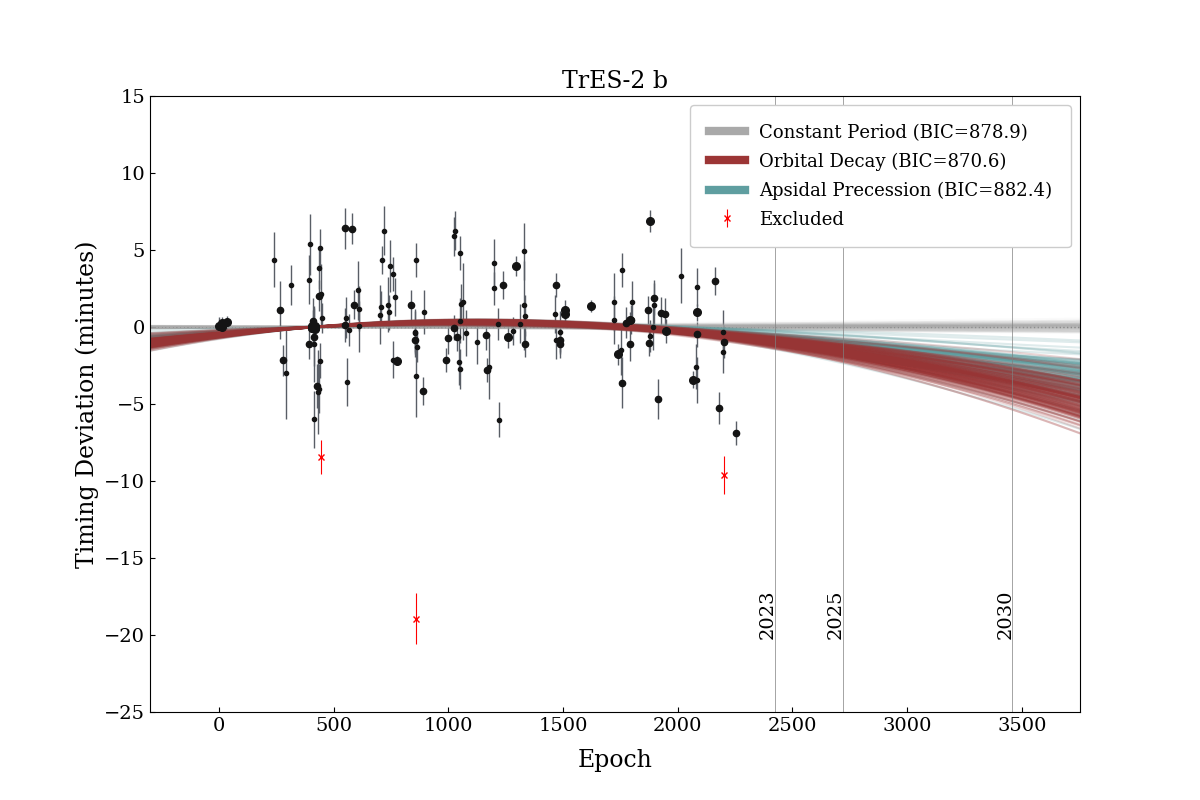}
\includegraphics[width=0.9\textwidth]{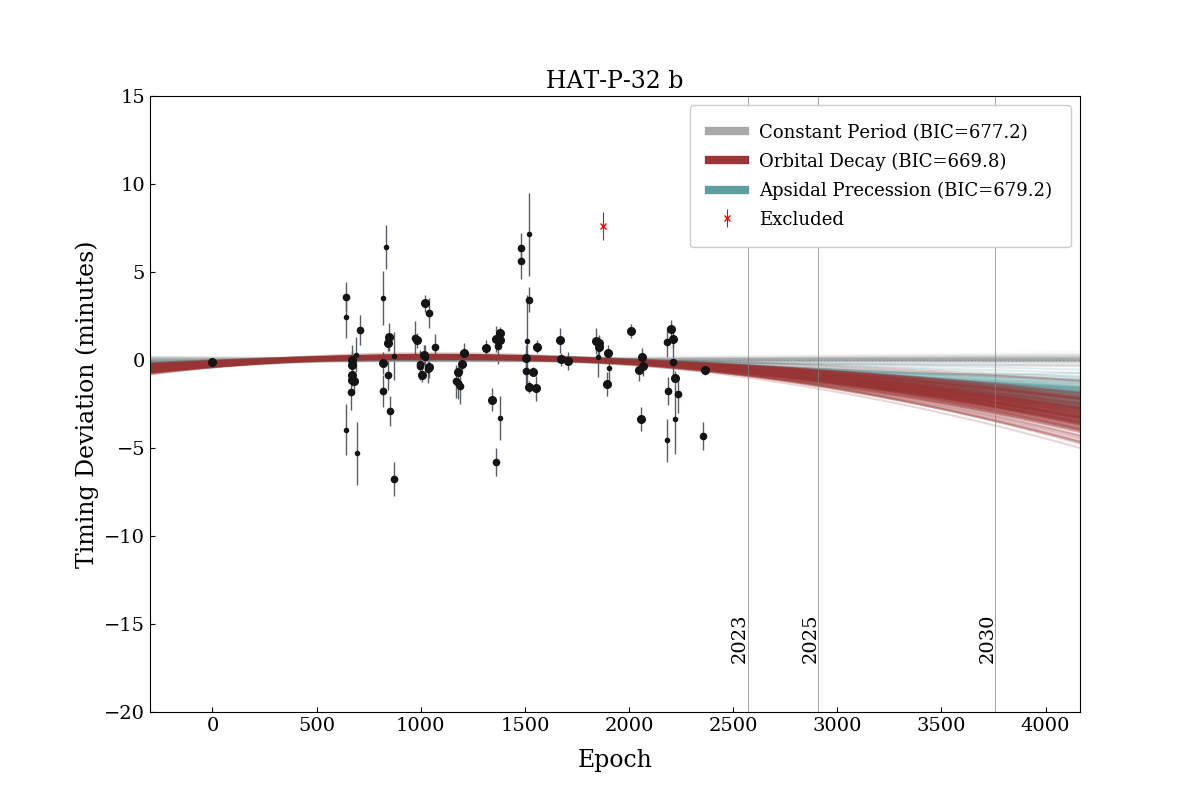}
\caption{Similar to Figure \ref{fig:wasp12wasp4}, but for TrES-2\,b (top) and HAT-P-32\,b (bottom).}
\label{fig:tres2hatp32}
\end{figure*}

TrES-2\,b is a $1.2$ M\textsubscript{J}, $1.2$ R\textsubscript{J} Hot Jupiter discovered in 2006 on a $2.470613374 \pm 0.000000019$ day, near grazing orbit \citep{odonovan_tres-2_2006,bonomo_gaps_2017}. As TrES-2\,b lies within the Kepler field, it was extensively studied in the years following its discovery. Early publications suggest evidence of short-term transit timing and duration variations, with explanations including a third body in the system or the existence of a moon \citep{rabus_transit_2009,mislis_detection_2009}. Later studies, however, were not able to support these claims, thus citing no evidence of short or long-term transit timing variations for TrES-2\,b  \citep{kipping_analysis_2011,schroter_consistent_2012,raetz_transit_2014}. We have not found any TTV studies of TrES-2\,b in years since.\par  

The initial analysis of the TrES-2\,b ETD data showed a strong preference for the orbital decay model with a period derivative of $-20.7 \pm 2.1$ ms/yr (Table \ref{table:table1}). The $\Delta\text{BIC}$ favours the decay model over a constant period by a significant value of $46.8$. Because of this, TrES-2\,b was selected as a target of interest for further analysis. The ETD hosts 331 transits of TrES-2\,b spanning the years 2006-2021, 270 of which are DQ 1-3. After removal of partial and duplicate transits 152 remain (Table \ref{table:tres2data}), as many were partial transits. The analysis of these cleaned data yields a less convincing, but still notable, argument for the decay scenario. It is worth highlighting that the first few epochs are mostly literature observations, which may be driving the model due to their generally higher precision and having been measured several years before the ETD observations begin (Figure \ref{fig:tres2hatp32}). The $\Delta\text{BIC}$ favours the decay model over a linear ephemeris by $8.3$, with the best-fit decay rate being $-12.6 \pm 2.4$ ms/yr (Table \ref{table:table2}).\par

Though the $\Delta\text{BIC}$ metric favours the orbital decay model over a constant period, it does not favour the apsidal precession model over a constant period. If there is statistically significant curvature in the timing residuals, it is reasonable to expect that the apsidal precession model should be a better fit than a constant period. Looking at only the $\chi^2$ values, the constant period model ($\chi^2=868.8$) is a worse fit than both apsidal precession ($\chi^2=857.4$) and orbital decay ($\chi^2=855.6$) models. The distinction between the two timing models is less clear in the case of TrES-2\,b than for the other systems explored here due to the large spread in the ETD data and the high number of DQ 3 observations (Figure \ref{fig:tres2hatp32}). We found that TrES-2\,b was the only target that could not pass the MCMC convergence criteria for the apsidal precession model. The resulting best-fit apsidal precession rate and eccentricity are $0.00061^{+0.00023}_{-0.00022}$ rad per epoch and $0.0030^{+0.0039}_{-0.0014}$, respectively (Table \ref{table:tres2results}). 

\subsubsection{HAT-P-32\,b}\label{sec:hatp32}

HAT-P-32\,b is a $0.8$ M\textsubscript{J}, $1.8$ R\textsubscript{J} Hot Jupiter discovered in 2011 on a $2.15000825 \pm 0.00000012$ day orbit \citep{hartman_hat-p-32b_2011,bonomo_gaps_2017}. In 2014 it was the subject of a TTV study seeking evidence of an additional body in the system, but amplitudes of greater than 1.5 minutes were ruled out \citep{seeliger_transit_2014}. The contributed observations on the ETD span the years 2007-2021, though there is a 633 orbit gap after the discovery transit. During a preliminary examination of the HAT-P-32\,b transit centres it was noted that, without the discovery epoch, there is visual curvature in the timing residuals.\footnote{\url{http://var2.astro.cz/ETD/etd.php?STARNAME=HAT-P-32&PLANET=b}} This point was removed for the initial round of analysis so that the significance of the curvature could be examined and compared with other systems. HAT-P-32\,b was selected for further analysis, as orbital decay at a rate of  $-30.2 \pm 3.3$ ms/yr is favoured over a constant period ephemeris by $\Delta\text{BIC}=39.7$ (Table \ref{table:table1}).\par

The ETD hosts 167 transits of HAT-P-32\,b, 142 of which are DQ 1-3. After the exclusion of partial and duplicate transits, 88 observations remain (Table \ref{table:hatp32data}). This includes the discovery epoch, which was added back to the data set as it was found that the transit time was not converted from BJD$_{\rm UTC}$ to HJD$_{\rm UTC}$ when listed on the database \citep{hartman_hat-p-32b_2011}. Results from the final analysis of the cleaned HAT-P-32\,b data maintain a preference for the orbital decay model over a constant period by a $\Delta\text{BIC}$ of $7.4$, as well as the apsidal precession model by a $\Delta\text{BIC}$ of $9.4$ (Figure \ref{fig:tres2hatp32}). However, the $\chi^2$ of the orbital decay and apsidal precession models are indistinguishable. The best-fit period derivative for the decay model is $-7.3 \pm 1.5$ ms/yr, and the apsidal precession model yields an eccentricity of $0.0022^{+0.0049}_{-0.0012}$ and precession rate $0.00054^{+0.00028}_{-0.00024}$ rad/epoch (Table \ref{table:hatp32results}).\par 

\subsubsection{TrES-3\,b}\label{sec:tres3}
\begin{figure*}
\includegraphics[width=0.9\textwidth]{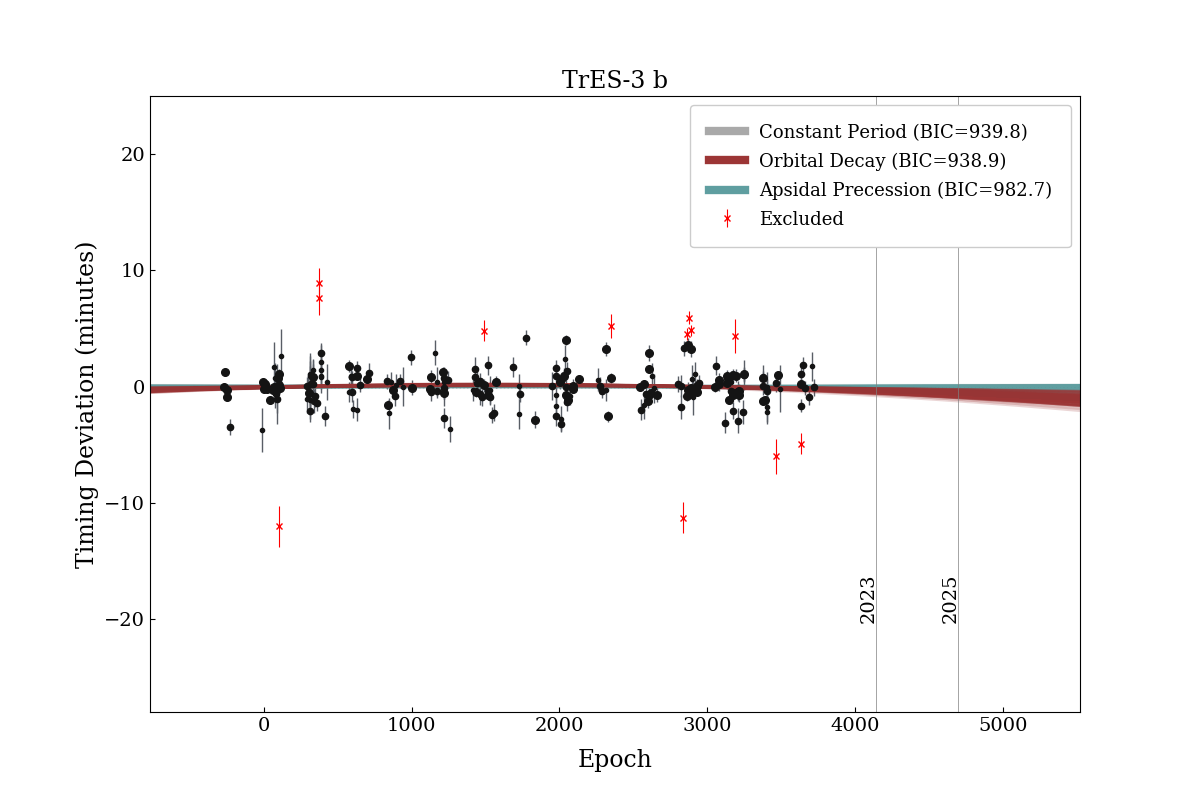}
\includegraphics[width=0.9\textwidth]{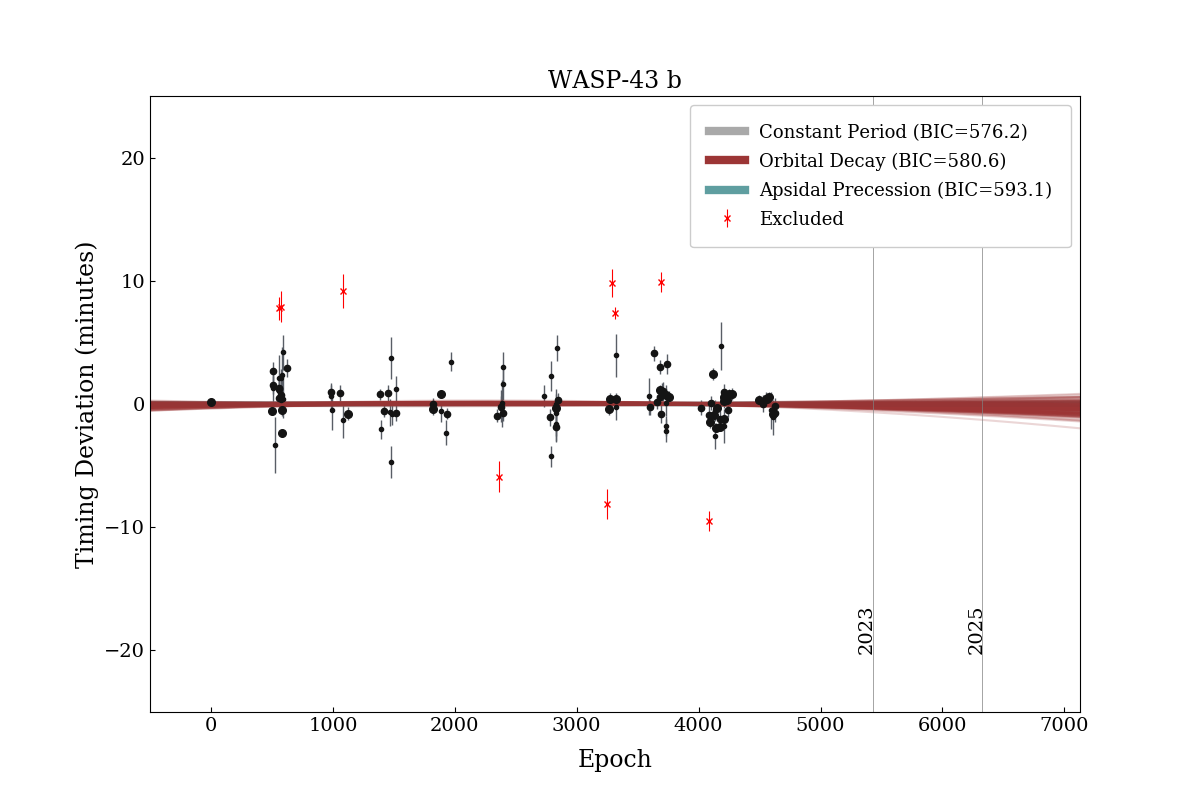}
\caption{Similar to Figure \ref{fig:wasp12wasp4}, but for TrES-3\,b (top) and WASP-43\,b (bottom).}
\label{fig:tres3wasp43}
\end{figure*}

TrES-3\,b is a $1.8$ M\textsubscript{J}, $1.3$ R\textsubscript{J} Hot Jupiter with an orbital period of $1.306186483 \pm 0.00000007$ days that was discovered in 2007 \citep{odonovan_tres-3_2007,bonomo_gaps_2017}. It has been extensively studied in the search for periodic TTVs indicating the presence of another planet, but no conclusive evidence of such variations have been found \citep{sozzetti_new_2009,gibson_transit_2009,jiang_possible_2013,vanko_photometric_2013,puskullu_photometric_2017}. \citet{zhao_photometric_2018} and \citet{mannaday_probing_2020} have investigated the possibility of orbital decay, both concluding that TrES-3\,b transit times are consistent with a constant period. However, \citet{mannaday_probing_2020} found the difference in BIC values of the orbital decay and constant period models to be very similar and recommend further observations.\par 

The ETD hosts 556 observations of TrES-3\,b from the years 2007-2021, 465 of which are DQ 1-3. This is one of the longest and most complete observational baselines in this study. In the initial round of analysis the orbital decay model was favoured for TrES-3\,b by a $\Delta\text{BIC}$ of $26.7$ when compared to the linear model (Table \ref{table:table1}). The best-fit decay rate is small at $-5.08 \pm 0.63$ ms/yr, but it was selected for further analysis. Many of the observations of TrES-3\,b on the ETD do not have a clear ingress or egress and many of the early observations do not have a recorded timing uncertainty, resulting in their immediate exclusion. After the removal of partial transits and duplicates (including the first literature observation) there are 231 observations remaining (Table \ref{table:tres3data}). The second, follow up analysis on this cleaned data set is more favourable toward a constant period model (Figure \ref{fig:tres3wasp43}). The best-fit decay rate is $-2.75 \pm 0.78$ ms/yr, consistent with a constant period when considering the variance in the data. The BIC values for the decay and constant period models are indistinguishable (Table \ref{table:table2}). The precession model is strongly disfavoured by its BIC, which can be explained by the apparent lack of curvature in the timing residuals. Indeed, the precession model is driven to a very low eccentricity of $0.0005$ and a slow precession rate of $0.00013$ rad per epoch, and is ultimately a poor fit (Table \ref{table:tres3results}).\par 

\subsubsection{WASP-43\,b}\label{sec:wasp43}

WASP-43\,b is a $2.1$ M\textsubscript{J}, $1.0$ R\textsubscript{J} Hot Jupiter on a 0.81347437 $\pm$ 0.00000013 day orbit \citep{hellier_wasp-43b_2011,bonomo_gaps_2017}. The WASP-43\,b system has been monitored since 2014 for signs of long-term transit timing variations because its ultra-short orbital period is thought to make it a good candidate for exhibiting orbital decay \citep{blecic_spitzer_2014,murgas_gtc_2014,chen_broad-band_2014,ricci_multi-filter_2014}. In 2016, \citet{jiang_possible_2016} detected evidence of orbital decay at a rate of $-0.02890795 \pm 0.00772547$ s/year. Later that same year, \citet{hoyer_ruling_2016} published a homogeneous analysis including new data that showed no indication of a period variation. Two years later, \citet{zhao_photometric_2018} found that orbital decay is slightly preferred over a constant period, but at a rate of $-0.005248 \pm 0.001714$ s/yr. In this study, the initial analysis of ETD transit centres strongly favours a constant period (Table \ref{table:table1}), but the system was selected for further analysis due to the interest in the literature.\par 

The ETD hosts 186 transits of WASP-43\,b spanning the years 2010-2021, with 162 being DQ 1-3. After the exclusion of partial and duplicate transits there are 126 remaining observations (Table \ref{table:wasp43data}). Analysis of the cleaned data supports the constant period model for WASP-43\,b over orbital decay by $\Delta\text{BIC}$ of $4.4$ (Figure \ref{fig:tres3wasp43}). In addition to this, the best-fit orbital decay rate is $-1.0 \pm 1.2$ ms/yr, which is consistent with a constant orbital period (Table \ref{table:table2}). Like TrES-3\,b, apsidal precession results in poor fit due to the sinusoidal nature of the model and the strong linearity of the data (Table \ref{table:wasp43results}).\par

%%% DISCUSSION %%%

\section{Discussion and Conclusions} \label{sec:discussion}

This study has utilized the wealth of citizen science observations on the Exoplanet Transit Database (ETD) to identify Hot Jupiter systems that are candidates for the detection of orbital decay, or more precisely, detectable period drifts. In the process, we have demonstrated the potential of citizen scientists to contribute to the detection of long-term orbital evolution signatures of close-in giant planets. \par 

We highlight systems in the ETD that should be prioritized for further study in Table \ref{table:table2}. We find that eight star-planet systems in the ETD show statistically compelling quadratic trends in the transit times and particularly recommend future follow-up of HAT-P-19\,b, HAT-P-32\,b, TrES-1\,b, and WASP-10\,b.\par

For an additional approach, we purposefully take a pessimistic view of the data's predictive power and assume that the uncertainties of the ETD transit centres are generally unreliable. In this case, we use the variance of the timing residuals during fitting and find that three of the targets (WASP-12\,b, HAT-P-19\,b, and TrES-1\,b) still show evidence of a negative period derivative. \par

In addition to orbital decay, we consider apsidal precession as a source of timing variations -- in no case is this model preferred over orbital decay. In many cases, however, the $\chi^2$ value of the orbital decay and apsidal precession models are similar. We further note that only the low-eccentricity ($e<<0.1$) expansion for precession is used in this work (as well as typically in others). Preliminary analysis of WASP-12\,b using the higher-order eccentricity expansion for precession, up to $e^5$ as presented in \citet{ragozzine_probing_2009}, does show promise in fitting the data, but only if a high eccentricity ($e\approx 0.1$) is used. The overall results are also not clearly better than the orbital decay model and so are not pursued further here. \par

Confirmation of period drifts, regardless of the cause, would provide additional constraints on the system. For orbital decay, the period derivative can be used to infer properties such as the planet's remaining lifetime and the stellar tidal dissipation rate. For precession, the model could constrain the interior density distribution of Hot Jupiters and potentially yield an upper limit on orbital eccentricities (of course, radial velocity data, if available, provide independent eccentricity constraints). Other possible effects, such as line-of-sight acceleration and stellar activity, will also need to be considered in understanding the underlying cause of observed timing deviations.\par

As our study aims to aid in selecting systems for long-term observing campaigns, in addition to extending the observed baselines of those systems, a next step is to conduct in-depth analyses into the identified candidates. Such studies should combine all available data sets for the system of interest, homogeneously refitting them where possible. We will aim to conduct such studies, also seeking to obtain further observations of these systems, particularly by collaborating with citizen scientists. \par

The exoplanet WASP-12\,b has proved to be critical as a control for this study and serves as an important test case for orbital decay searches in general. The consistency between the results using citizen science data and those in the literature demonstrates the feasibility of such work. The ETD offers a rich history of transit times for hundreds of star-planet systems that can make the search for evidence of long-term dynamical evolution, orbital decay in this case, more efficient. While this work does not suggest that any of the identified systems, apart from WASP-12\,b, are indeed undergoing orbital decay or other period deviations, the work does suggest that there is sufficient evidence to keep looking.\par

%%% ACKNOWLEDGMENTS %%%

\section*{Acknowledgements}

This work would not be possible without the dedication and skill of the Exoplanet Transit Database (ETD) coordinators and observers, and we sincerely thank them all for their ongoing contributions. The names of observers of the data used for this study - as given on ETD with minor formatting changes - can be found in the supplementary data tables on Zenodo at \url{https://doi.org/10.5281/zenodo.7098460} and in Appendix \ref{sec:appendixB}. As of May 13, 2022 there are 625 ETD contributors consisting of universities, teams, and individuals from around the world. Additionally, this network of observers expands with the inclusion of the ExoClock and Exoplanet Watch programmes. We emphasize the science potential of such a large community of ground-based observers, and encourage continued collaboration and the inclusion of new observers. Having identified key systems to follow-up in further depth, we look forward to collaborating with the participants of these programmes as we continue to investigate the potential for non-linear ephemerides. \par

We thank Darin Ragozzine for helpful comments on the draft of this paper. This work is supported, in part, by an NSERC Discovery Grant, an NSERC CREATE grant, and the Canada Research Chairs program. BE is a Laureate of the Paris Region fellowship programme which is supported by the Ile-de-France Region and has received funding under the Horizon 2020 innovation framework programme and the Marie Sklodowska-Curie grant agreement no. 945298.  \par

%%% BIBLIOGRAPHY %%%
\newpage
\bibliography{ETDpaper}{}

\begin{thebibliography}{}
\expandafter\ifx\csname natexlab\endcsname\relax\def\natexlab#1{#1}\fi
\providecommand{\url}[1]{\href{#1}{#1}}
\providecommand{\dodoi}[1]{doi:~\href{http://doi.org/#1}{\nolinkurl{#1}}}
\providecommand{\doeprint}[1]{\href{http://ascl.net/#1}{\nolinkurl{http://ascl.net/#1}}}
\providecommand{\doarXiv}[1]{\href{https://arxiv.org/abs/#1}{\nolinkurl{https://arxiv.org/abs/#1}}}

\bibitem[{Alonso {et~al.}(2004)Alonso, Brown, Torres, Latham, Sozzetti,
  Mandushev, Belmonte, Charbonneau, Deeg, Dunham, O’Donovan, \&
  Stefanik}]{alonso_tres-1_2004}
Alonso, R., Brown, T.~M., Torres, G., {et~al.} 2004, The Astrophysical Journal,
  613, L153, \dodoi{10.1086/425256}

\bibitem[{{Ba{\c{s}}t{\"u}rk} {et~al.}(2020){Ba{\c{s}}t{\"u}rk},
  {Yal{\c{c}}{\i}nkaya}, {Esmer}, {Tanr{\i}verdi}, {Mancini}, {Daylan},
  {Southworth}, \& {Keten}}]{basturk_holistic_2020}
{Ba{\c{s}}t{\"u}rk}, {\"O}., {Yal{\c{c}}{\i}nkaya}, S., {Esmer}, E.~M.,
  {et~al.} 2020, \mnras, 496, 4174, \dodoi{10.1093/mnras/staa1758}

\bibitem[{Baluev {et~al.}(2015)Baluev, Sokov, Shaidulin, Sokova, Jones, Tuomi,
  Anglada-Escudé, Benni, Colazo, Schneiter, D'Angelo, Burdanov,
  Fernández-Lajús, Baştürk, Hentunen, \&
  Shadick}]{baluev_benchmarking_2015}
Baluev, R.~V., Sokov, E.~N., Shaidulin, V.~S., {et~al.} 2015, MNRAS, 450, 3101,
  \dodoi{10.1093/mnras/stv788}

\bibitem[{Baluev {et~al.}(2019)Baluev, Sokov, Jones, Shaidulin, Sokova,
  Nielsen, Benni, Schneiter, Villarreal D’Angelo, Fernández-Lajús,
  Di Sisto, Baştürk, Bretton, Wunsche, Hentunen, Shadick, Jongen, Kang, Kim,
  Pakštienė, Qvam, Knight, Guerra, Marchini, Salvaggio, Papini, Evans,
  Salisbury, Garcia, Molina, Garlitz, Esseiva, Ogmen, Karavaev, Rusov,
  Ibrahimov, \& Karimov}]{baluev_homogeneously_2019}
Baluev, R.~V., Sokov, E.~N., Jones, H. R.~A., {et~al.} 2019, MNRAS, 490, 1294,
  \dodoi{10.1093/mnras/stz2620}

\bibitem[{Baluev {et~al.}(2020)Baluev, Sokov, Hoyer, Huitson, da Silva, Evans,
  Sokova, Knight, \& Shaidulin}]{baluev_wasp-4_2020}
Baluev, R.~V., Sokov, E.~N., Hoyer, S., {et~al.} 2020, MNRAS. Letters, 496,
  L11, \dodoi{10.1093/mnrasl/slaa069}

\bibitem[{Barros {et~al.}(2013)Barros, Boué, Gibson, Pollacco, Santerne,
  Keenan, Skillen, \& Street}]{barros_transit_2013}
Barros, S. C.~C., Boué, G., Gibson, N.~P., {et~al.} 2013, MNRAS, 430, 3032,
  \dodoi{10.1093/mnras/stt111}

\bibitem[{Batygin {et~al.}(2016)Batygin, Bodenheimer, \&
  Laughlin}]{batygin_situ_2016}
Batygin, K., Bodenheimer, P.~H., \& Laughlin, G.~P. 2016, The Astrophysical
  Journal, 829, 114, \dodoi{10.3847/0004-637X/829/2/114}

\bibitem[{Birkby {et~al.}(2014)Birkby, Cappetta, Cruz, Koppenhoefer, Ivanyuk,
  Mustill, Hodgkin, Pinfield, Sipőcz, Kovács, Saglia, Pavlenko, Barrado,
  Bayo, Campbell, Catalan, Fossati, Gálvez-Ortiz, Kenworthy, Lillo-Box,
  Martín, Mislis, de~Mooij, Nefs, Snellen, Stoev, Zendejas, Burgo, Barnes,
  Goulding, Haswell, Kuznetsov, Lodieu, Murgas, Palle, Solano, Steele, \&
  Tata}]{birkby_wts-2_2014}
Birkby, J.~L., Cappetta, M., Cruz, P., {et~al.} 2014, MNRAS, 440, 1470,
  \dodoi{10.1093/mnras/stu343}

\bibitem[{Blecic {et~al.}(2014)Blecic, Harrington, Madhusudhan, Stevenson,
  Hardy, Cubillos, Hardin, Bowman, Nymeyer, Anderson, Hellier, Smith, \&
  Collier~Cameron}]{blecic_spitzer_2014}
Blecic, J., Harrington, J., Madhusudhan, N., {et~al.} 2014, The Astrophysical
  Journal, 781, 116, \dodoi{10.1088/0004-637X/781/2/116}

\bibitem[{Boley {et~al.}(2016)Boley, Granados~Contreras, \&
  Gladman}]{boley_situ_2016}
Boley, A.~C., Granados~Contreras, A.~P., \& Gladman, B. 2016, Astrophysical
  journal. Letters, 817, L17, \dodoi{10.3847/2041-8205/817/2/L17}

\bibitem[{Bonomo {et~al.}(2017)Bonomo, Desidera, Benatti, Borsa, Crespi,
  Damasso, Lanza, Sozzetti, Lodato, Marzari, Boccato, Claudi, Cosentino,
  Covino, Gratton, Maggio, Micela, Molinari, Pagano, Piotto, Poretti,
  Smareglia, Affer, Biazzo, Bignamini, Esposito, Giacobbe, Hébrard, Malavolta,
  Maldonado, Mancini, Martinez~Fiorenzano, Masiero, Nascimbeni, Pedani, Rainer,
  \& Scandariato}]{bonomo_gaps_2017}
Bonomo, A.~S., Desidera, S., Benatti, S., {et~al.} 2017, Astronomy and
  Astrophysics, 602, A107, \dodoi{10.1051/0004-6361/201629882}

\bibitem[{{Bouma} {et~al.}(2020){Bouma}, {Winn}, {Howard}, {Howell},
  {Isaacson}, {Knutson}, \& {Matson}}]{bouma_etal_2020}
{Bouma}, L.~G., {Winn}, J.~N., {Howard}, A.~W., {et~al.} 2020, ApJ, 893, L29,
  \dodoi{10.3847/2041-8213/ab8563}

\bibitem[{Bouma {et~al.}(2020)Bouma, Winn, Howard, Howell, Isaacson, Knutson,
  \& Matson}]{bouma_wasp-4_2020}
Bouma, L.~G., Winn, J.~N., Howard, A.~W., {et~al.} 2020, Astrophysical journal.
  Letters, 893, L29, \dodoi{10.3847/2041-8213/ab8563}

\bibitem[{Bouma {et~al.}(2019)Bouma, Winn, Baxter, Bhatti, Dai, Daylan,
  Désert, Hill, Kane, Stassun, Villasenor, Ricker, Vanderspek, Latham, Seager,
  Jenkins, Berta-Thompson, Colón, Fausnaugh, Glidden, Guerrero, Rodriguez,
  Twicken, \& Wohler}]{bouma_wasp-4b_2019}
Bouma, L.~G., Winn, J.~N., Baxter, C., {et~al.} 2019, The Astronomical Journal,
  157, 217, \dodoi{10.3847/1538-3881/ab189f}

\bibitem[{Chen {et~al.}(2014)Chen, van Boekel, Wang, Nikolov, Fortney, Seemann,
  Wang, Mancini, \& Henning}]{chen_broad-band_2014}
Chen, G., van Boekel, R., Wang, H., {et~al.} 2014, Astronomy and Astrophysics,
  563, A40, \dodoi{10.1051/0004-6361/201322740}

\bibitem[{Christian {et~al.}(2009)Christian, Gibson, Simpson, Street, Skillen,
  Pollacco, Collier~Cameron, Joshi, Keenan, Stempels, Haswell, Horne, Anderson,
  Bentley, Bouchy, Clarkson, Enoch, Hebb, Hébrard, Hellier, Irwin, Kane,
  Lister, Loeillet, Maxted, Mayor, McDonald, Moutou, Norton, Parley, Pont,
  Queloz, Ryans, Smalley, Smith, Todd, Udry, West, Wheatley, \&
  Wilson}]{christian_wasp-10b_2009}
Christian, D.~J., Gibson, N.~P., Simpson, E.~K., {et~al.} 2009, MNRAS, 392,
  1585, \dodoi{10.1111/j.1365-2966.2008.14164.x}

\bibitem[{Collins(2019)}]{collins_tess_2019}
Collins, K. 2019, in American {Astronomical} {Society} {Meeting} {Abstracts},
  Vol. 233, American {Astronomical} {Society} {Meeting} {Abstracts} \#233,
  140.05

\bibitem[{Davoudi {et~al.}(2021)Davoudi, MirshafieKhozani, Paki, Roshana,
  Hasheminasab, MazidabadiFarahani, Ahangarani~Farahani, Farjadnia,
  Nasrollahzadeh, Rezvanpanah, Mousavi, Foroughi, Poro, \&
  Ghalee}]{davoudi_refined_2021}
Davoudi, F., MirshafieKhozani, P., Paki, E., {et~al.} 2021, Astronomy Letters,
  47, 638, \dodoi{10.1134/S1063773721090024}

\bibitem[{Eastman {et~al.}(2010)Eastman, Siverd, \&
  Gaudi}]{eastman_achieving_2010}
Eastman, J., Siverd, R., \& Gaudi, B.~S. 2010, Publications of the Astronomical
  Society of the Pacific, 122, 935, \dodoi{10.1086/655938}

\bibitem[{{Edwards} {et~al.}(2020){Edwards}, {Anisman}, {Changeat}, {Morvan},
  {Wright}, {Yip}, {Abdullahi}, {Ali}, {Amofa}, {Antoniou}, {Arzouni},
  {Bradley}, {Campana}, {Chavda}, {Creswell}, {Gazieva}, {Gudgeon-Sidelnikova},
  {Guha}, {Hayden}, {Huda}, {Hussein}, {Ibrahim}, {Ike}, {Jama}, {Joshi}, {Kc},
  {Keenan}, {Kelly-Smith}, {Khan}, {Korodimos}, {Liang}, {Nogueira},
  {Martey-Botchway}, {Masruri}, {Miyamaru}, {Moalin}, {Monteiro}, {Nawrocka},
  {Musa}, {Nelson}, {Ogunjuyigbe}, {Patel}, {Pereira}, {Ramsey}, {Rasoul},
  {Reetsong}, {Saeed}, {Sander}, {Sanetra}, {Tarabe}, {Tareke}, {Tasneem},
  {Teo}, {Uddin}, {Upadhyay}, {Yanakiev}, {Yatingiri}, {Dunn}, {Kokori},
  {Tsiaras}, {Gomez}, {Tinetti}, \& {Tennyson}}]{edwards_orbytsI}
{Edwards}, B., {Anisman}, L., {Changeat}, Q., {et~al.} 2020, Research Notes of
  the AAS, 4, 109, \dodoi{10.3847/2515-5172/aba42b}

\bibitem[{{Edwards} {et~al.}(2021){Edwards}, {Ho}, {Osborne}, {Deen},
  {Hathorn}, {Johnson}, {Patel}, {Vogireddy}, {Waddon}, {Ahmed}, {Bham},
  {Campbell}, {Chummun}, {Crossley}, {Dunsdon}, {Hayes}, {Malik}, {Marsden},
  {Mayfield}, {Mitchell}, {Prosser}, {Rabrenovic}, {Smith}, {Thomas}, {Kokori},
  {Tsiaras}, {Tessenyi}, {Tinetti}, \& {Tennyson}}]{edwards_orbytsII}
{Edwards}, B., {Ho}, C. S.~K., {Osborne}, H. L.~M., {et~al.} 2021, ATOM, 2, 25,
  \dodoi{10.32374/atom.2020.2.4}

\bibitem[{Edwards {et~al.}(2021)Edwards, Changeat, Yip, Tsiaras, Taylor,
  Akhtar, AlDaghir, Bhattarai, Bhudia, Chapagai, Huang, Kabir, Khag, Khaliq,
  Khatri, Kneth, Kothari, Najmudin, Panchalingam, Patel, Premachandran, Qayyum,
  Rana, Shaikh, Syed, Theti, Zaidani, Saraf, de Mijolla, Caines, Kokori,
  Rocchetto, Mallonn, Bachschmidt, Bosch, Bretton, Chatelain, Deldem,
  Di Sisto, Evans, Fernández-Lajús, Guerra, Grau Horta, Kang, Kim, Leroy,
  Lomoz, de Haro, Hentunen, Jongen, Molina, Montaigut, Naves, Raetz, Sauer,
  Watkins, Wünsche, Zibar, Dunn, Tessenyi, Savini, Tinetti, \&
  Tennyson}]{edwards_original_2021}
Edwards, B., Changeat, Q., Yip, K.~H., {et~al.} 2021, MNRAS, 504, 5671,
  \dodoi{10.1093/mnras/staa1245}

\bibitem[{{Fabrycky} \& {Tremaine}(2007)}]{fabrycky_HJ_tides}
{Fabrycky}, D., \& {Tremaine}, S. 2007, ApJ, 669, 1298, \dodoi{10.1086/521702}

\bibitem[{Ford(2006)}]{ford_improving_2006}
Ford, E.~B. 2006, The Astrophysical Journal, 642, 505, \dodoi{10.1086/500802}

\bibitem[{Foreman-Mackey {et~al.}(2013)Foreman-Mackey, Hogg, Lang, \&
  Goodman}]{foreman-mackey_emcee_2013}
Foreman-Mackey, D., Hogg, D.~W., Lang, D., \& Goodman, J. 2013, Publications of
  the Astronomical Society of the Pacific, 125, 306, \dodoi{10.1086/670067}

\bibitem[{Gibson {et~al.}(2009)Gibson, Pollacco, Simpson, Barros, Joshi, Todd,
  Keenan, Skillen, Benn, Christian, Hrudková, \& Steele}]{gibson_transit_2009}
Gibson, N.~P., Pollacco, D., Simpson, E.~K., {et~al.} 2009, The Astrophysical
  Journal, 700, 1078, \dodoi{10.1088/0004-637X/700/2/1078}

\bibitem[{{Hartman} {et~al.}(2011{\natexlab{a}}){Hartman}, {Bakos}, {Sato},
  {Torres}, {Noyes}, {Latham}, {Kov{\'a}cs}, {Fischer}, {Howard}, {Johnson},
  {Marcy}, {Buchhave}, {F{\"u}resz}, {Perumpilly}, {B{\'e}ky}, {Stefanik},
  {Sasselov}, {Esquerdo}, {Everett}, {Csubry}, {L{\'a}z{\'a}r}, {Papp}, \&
  {S{\'a}ri}}]{hartman_hat-p-18b_2011}
{Hartman}, J.~D., {Bakos}, G.~{\'A}., {Sato}, B., {et~al.} 2011{\natexlab{a}},
  ApJ, 726, 52, \dodoi{10.1088/0004-637X/726/1/52}

\bibitem[{{Hartman} {et~al.}(2011{\natexlab{b}}){Hartman}, {Bakos}, {Torres},
  {Latham}, {Kov{\'a}cs}, {B{\'e}ky}, {Quinn}, {Mazeh}, {Shporer}, {Marcy},
  {Howard}, {Fischer}, {Johnson}, {Esquerdo}, {Noyes}, {Sasselov}, {Stefanik},
  {Fernandez}, {Szklen{\'a}r}, {L{\'a}z{\'a}r}, {Papp}, \&
  {S{\'a}ri}}]{hartman_hat-p-32b_2011}
{Hartman}, J.~D., {Bakos}, G.~{\'A}., {Torres}, G., {et~al.}
  2011{\natexlab{b}}, ApJ, 742, 59, \dodoi{10.1088/0004-637X/742/1/59}

\bibitem[{Hebb {et~al.}(2008)Hebb, Collier-Cameron, Loeillet, Pollacco,
  Hébrard, Street, Bouchy, Stempels, Moutou, Simpson, Udry, Yoshi, West,
  Skillen, Wilson, McDonald, Gibson, \& Consortium}]{hebb_wasp-12b_2008}
Hebb, L., Collier-Cameron, A., Loeillet, B., {et~al.} 2008,
  \dodoi{10.1088/0004-637X/693/2/1920}

\bibitem[{Hellier {et~al.}(2011)Hellier, Anderson, Collier~Cameron, Gillon,
  Jehin, Lendl, Maxted, Pepe, Pollacco, Queloz, Ségransan, Smalley, Smith,
  Southworth, Triaud, Udry, \& West}]{hellier_wasp-43b_2011}
Hellier, C., Anderson, D.~R., Collier~Cameron, A., {et~al.} 2011, Astronomy and
  Astrophysics, 535, L7, \dodoi{10.1051/0004-6361/201117081}

\bibitem[{Hoyer {et~al.}(2016)Hoyer, Pallé, Dragomir, \&
  Murgas}]{hoyer_ruling_2016}
Hoyer, S., Pallé, E., Dragomir, D., \& Murgas, F. 2016, The Astronomical
  Journal, 151, 137, \dodoi{10.3847/0004-6256/151/6/137}

\bibitem[{Hrudková {et~al.}(2008)Hrudková, Skillen, Benn, Pollacco, Gibson,
  Joshi, Harmanec, \& Tulloch}]{hrudkova_searching_2008}
Hrudková, M., Skillen, I., Benn, C., {et~al.} 2008, Proceedings of the
  International Astronomical Union, 4, 446, \dodoi{10.1017/S1743921308026896}

\bibitem[{Ioannidis {et~al.}(2016)Ioannidis, Huber, \&
  Schmitt}]{ioannidis_p_how_2016}
Ioannidis, P., Huber, K.~F., \& Schmitt, J. 2016, Astronomy and Astrophysics,
  585, A72, \dodoi{10.1051/0004-6361/201527184}

\bibitem[{Ivshina \& Winn(2022)}]{ivshina_tess_2022}
Ivshina, E.~S., \& Winn, J.~N. 2022, The Astrophysical Journal Supplement
  Series, 259, 62, \dodoi{10.3847/1538-4365/ac545b}

\bibitem[{Jiang {et~al.}(2016)Jiang, Lai, Savushkin, Mkrtichian, Antonyuk,
  Griv, Hsieh, \& Yeh}]{jiang_possible_2016}
Jiang, I.-G., Lai, C.-Y., Savushkin, A., {et~al.} 2016, The Astronomical
  Journal, 151, 17, \dodoi{10.3847/0004-6256/151/1/17}

\bibitem[{Jiang {et~al.}(2013)Jiang, Yeh, Thakur, Wu, Chien, Lin, Chen, Hu,
  Sun, \& Ji}]{jiang_possible_2013}
Jiang, I.-G., Yeh, L.-C., Thakur, P., {et~al.} 2013, The Astronomical Journal,
  145, 68, \dodoi{10.1088/0004-6256/145/3/68}

\bibitem[{Johnson {et~al.}(2010)Johnson, Winn, Cabrera, \&
  Carter}]{johnson_erratum_2010}
Johnson, J., Winn, J., Cabrera, N., \& Carter, J. 2010, The Astrophysical
  Journal, 712, L122, \dodoi{10.1088/2041-8205/712/1/L122}

\bibitem[{Johnson {et~al.}(2009)Johnson, Winn, Cabrera, \&
  Carter}]{johnson_smaller_2009}
Johnson, J.~A., Winn, J.~N., Cabrera, N.~E., \& Carter, J.~A. 2009, The
  Astrophysical Journal, 692, L100, \dodoi{10.1088/0004-637X/692/2/L100}

\bibitem[{Kipping \& Bakos(2011)}]{kipping_analysis_2011}
Kipping, D., \& Bakos, G. 2011, The Astrophysical Journal, 733, 36,
  \dodoi{10.1088/0004-637X/733/1/36}

\bibitem[{{Kokori} {et~al.}(2022){Kokori}, {Tsiaras}, {Edwards}, {Rocchetto},
  \& {Tinetti}}]{kokori_2022}
{Kokori}, A., {Tsiaras}, A., {Edwards}, B., {Rocchetto}, M., \& {Tinetti}, G.
  2022, The Astrophysical Journal Supplement Series, 258, 40,
  \dodoi{10.3847/1538-4365/ac3a10}

\bibitem[{{Kokori} {et~al.}(2021){Kokori}, {Tsiaras}, {Edwards}, {Rocchetto},
  {Tinetti}, {W{\"u}nsche}, {Paschalis}, {Agnihotri}, {Bachschmidt}, {Bretton},
  {Caines}, {Cal{\'o}}, {Casali}, {Crow}, {Dawes}, {Deldem},
  {Deligeorgopoulos}, {Dymock}, {Evans}, {Falco}, {Ferratfiat}, {Fowler},
  {Futcher}, {Guerra}, {Hurter}, {Jones}, {Kang}, {Kim}, {Lee}, {Lopresti},
  {Marino}, {Mallonn}, {Mortari}, {Morvan}, {Mugnai}, {Nastasi}, {Perroud},
  {Pereira}, {Phillips}, {Pintr}, {Raetz}, {Regembal}, {Savage}, {Sedita},
  {Sioulas}, {Strikis}, {Thurston}, {Tomacelli}, \& {Tomatis}}]{kokori_2021}
{Kokori}, A., {Tsiaras}, A., {Edwards}, B., {et~al.} 2021, Experimental
  Astronomy, 53, 547, \dodoi{10.1007/s10686-020-09696-3}

\bibitem[{Levrard {et~al.}(2009)Levrard, Winisdoerffer, \&
  Chabrier}]{levrard_falling_2009}
Levrard, B., Winisdoerffer, C., \& Chabrier, G. 2009, The Astrophysical
  Journal, 692, L9, \dodoi{10.1088/0004-637X/692/1/L9}

\bibitem[{Liddle(2007)}]{liddle_information_2007}
Liddle, A.~R. 2007, MNRAS. Letters, 377, L74,
  \dodoi{10.1111/j.1745-3933.2007.00306.x}

\bibitem[{Lin {et~al.}(1996)Lin, Bodenheimer, \& Richardson}]{lin_orbital_1996}
Lin, D. N.~C., Bodenheimer, P., \& Richardson, D.~C. 1996, Nature (London),
  380, 606, \dodoi{10.1038/380606a0}

\bibitem[{Maciejewski {et~al.}(2021)Maciejewski, Fernández, Aceituno, Ramos,
  Dimitrov, Donchev, \& Ohlert}]{maciejewski_revisiting_2021}
Maciejewski, G., Fernández, M., Aceituno, F., {et~al.} 2021, Astronomy and
  Astrophysics, 656, A88, \dodoi{10.1051/0004-6361/202142424}

\bibitem[{Maciejewski {et~al.}(2011{\natexlab{a}})Maciejewski, Raetz,
  Nettelmann, Seeliger, Adam, Nowak, \& Neuhäuser}]{maciejewski_analysis_2011}
Maciejewski, G., Raetz, S., Nettelmann, N., {et~al.} 2011{\natexlab{a}},
  Astronomy and Astrophysics, 535, A7, \dodoi{10.1051/0004-6361/201117127}

\bibitem[{Maciejewski {et~al.}(2018)Maciejewski, Stangret, Ohlert, Basaran,
  Maciejczak, Puciata-Mroczynska, \& Boulanger}]{maciejewski_new_2018}
Maciejewski, G., Stangret, M., Ohlert, J., {et~al.} 2018, Information bulletin
  on variable stars, \dodoi{10.22444/IBVS.6243}

\bibitem[{Maciejewski {et~al.}(2011{\natexlab{b}})Maciejewski, Dimitrov,
  Neuhäuser, Tetzlaff, Niedzielski, Raetz, Chen, Walter, Marka, Baar,
  Krejcová, Budaj, Krushevska, Tachihara, Takahashi, \&
  Mugrauer}]{maciejewski_transit_2011}
Maciejewski, G., Dimitrov, D., Neuhäuser, R., {et~al.} 2011{\natexlab{b}},
  MNRAS, 411, 1204, \dodoi{10.1111/j.1365-2966.2010.17753.x}

\bibitem[{Maciejewski {et~al.}(2016)Maciejewski, Dimitrov, Fernández, Sota,
  Nowak, Ohlert, Nikolov, Bukowiecki, Hinse, Pallé, Tingley, Kjurkchieva, Lee,
  \& Lee}]{maciejewski_departure_2016}
Maciejewski, G., Dimitrov, D., Fernández, M., {et~al.} 2016, Astronomy and
  Astrophysics, 588, L6, \dodoi{10.1051/0004-6361/201628312}

\bibitem[{Mallonn {et~al.}(2019)Mallonn, von Essen, Herrero, Alexoudi, Granzer,
  Sosa, Strassmeier, Bakos, Bayliss, Brahm, Bretton, Campos, Carone, Colón,
  Dale, Dragomir, Espinoza, Evans, Garcia, Gu, Guerra, Jongen, Jordán, Kang,
  Keles, Kim, Lendl, Molina, Salisbury, Scaggiante, Shporer, Siverd, Sokov,
  Sokova, \& Wünsche}]{mallonn_ephemeris_2019}
Mallonn, M., von Essen, C., Herrero, E., {et~al.} 2019, Astronomy and
  Astrophysics, 622, A81, \dodoi{10.1051/0004-6361/201834194}

\bibitem[{Mandushev {et~al.}(2011)Mandushev, Dunham, Quinn, \&
  Latham}]{mandushev_tres-5_2011}
Mandushev, G., Dunham, E.~W., Quinn, S.~N., \& Latham, D.~W. 2011, The
  Astrophysical Journal, 741

\bibitem[{Mannaday {et~al.}(2020)Mannaday, Thakur, Jiang, Sahu, Joshi, Pandey,
  Joshi, Yadav, Su, Sariya, Yeh, Griv, Mkrtichian, Shlyapnikov, Moskvin,
  Ignatov, Va~ko, \& Püsküllü}]{mannaday_probing_2020}
Mannaday, V.~K., Thakur, P., Jiang, I.-G., {et~al.} 2020, The Astronomical
  Journal, 160, 47, \dodoi{10.3847/1538-3881/ab9818}

\bibitem[{Matsumura {et~al.}(2010)Matsumura, Peale, \&
  Rasio}]{matsumura_tidal_2010}
Matsumura, S., Peale, S.~J., \& Rasio, F.~A. 2010, The Astrophysical Journal,
  725, 1995, \dodoi{10.1088/0004-637X/725/2/1995}

\bibitem[{Mislis \& Schmitt(2009)}]{mislis_detection_2009}
Mislis, D., \& Schmitt, J. H. M.~M. 2009, Astronomy and Astrophysics, 500, L45,
  \dodoi{10.1051/0004-6361/200811424}

\bibitem[{Murgas {et~al.}(2014)Murgas, Pallé, Zapatero~Osorio, Nortmann,
  Hoyer, \& Cabrera-Lavers}]{murgas_gtc_2014}
Murgas, F., Pallé, E., Zapatero~Osorio, M.~R., {et~al.} 2014, Astronomy and
  Astrophysics, 563, A41, \dodoi{10.1051/0004-6361/201322374}

\bibitem[{{O'Donovan} {et~al.}(2007){O'Donovan}, {Charbonneau}, {Bakos},
  {Mandushev}, {Dunham}, {Brown}, {Latham}, {Torres}, {Sozzetti}, {Kov{\'a}cs},
  {Everett}, {Baliber}, {Hidas}, {Esquerdo}, {Rabus}, {Deeg}, {Belmonte},
  {Hillenbrand}, \& {Stefanik}}]{odonovan_tres-3_2007}
{O'Donovan}, F.~T., {Charbonneau}, D., {Bakos}, G.~{\'A}., {et~al.} 2007, ApJ,
  663, L37, \dodoi{10.1086/519793}

\bibitem[{O’Donovan {et~al.}(2006)O’Donovan, Charbonneau, Mandushev,
  Dunham, Latham, Torres, Sozzetti, Brown, Trauger, Belmonte, Rabus, Almenara,
  Alonso, Deeg, Esquerdo, Falco, Hillenbrand, Roussanova, Stefanik, \&
  Winn}]{odonovan_tres-2_2006}
O’Donovan, F.~T., Charbonneau, D., Mandushev, G., {et~al.} 2006, The
  Astrophysical Journal, 651, L61, \dodoi{10.1086/509123}

\bibitem[{Patra {et~al.}(2020)Patra, Winn, Holman, Gillon, \&
  Burdanov}]{patra_continuing_2020}
Patra, K.~C., Winn, J.~N., Holman, M.~J., Gillon, M., \& Burdanov, A. 2020, The
  Astronomical Journal, 159, 150, \dodoi{10.3847/1538-3881/ab7374}

\bibitem[{Patra {et~al.}(2017)Patra, Winn, Holman, Yu, Deming, \&
  Dai}]{patra_apparently_2017}
Patra, K.~C., Winn, J.~N., Holman, M.~J., {et~al.} 2017, The Astronomical
  Journal, 154, 4, \dodoi{10.3847/1538-3881/aa6d75}

\bibitem[{Penev {et~al.}(2018)Penev, Bouma, Winn, \&
  Hartman}]{penev_empirical_2018}
Penev, K., Bouma, L.~G., Winn, J.~N., \& Hartman, J.~D. 2018, The Astronomical
  Journal, 155, 165, \dodoi{10.3847/1538-3881/aaaf71}

\bibitem[{Petrucci {et~al.}(2019)Petrucci, Jofré, Gómez Maqueo~Chew, Hinse,
  Mašek, Tan, \& Gómez}]{petrucci_discarding_2019}
Petrucci, R., Jofré, E., Gómez Maqueo~Chew, Y., {et~al.} 2019, MNRAS,
  \dodoi{10.1093/mnras/stz3034}

\bibitem[{Petrucci {et~al.}(2013)Petrucci, Jofré, Schwartz, Cúneo, Martínez,
  Gómez, Buccino, \& Mauas}]{petrucci_no_2013}
Petrucci, R., Jofré, E., Schwartz, M., {et~al.} 2013, Astrophysical journal.
  Letters, 779, L23

\bibitem[{Poddaný {et~al.}(2011)Poddaný, Brat, \& Pejcha}]{poddany_new_2011}
Poddaný, S., Brat, L., \& Pejcha, O. 2011, EPJ Web of Conferences, 11, 6008,
  \dodoi{10.1051/epjconf/20101106008}

\bibitem[{Poddaný {et~al.}(2010)Poddaný, Brát, \&
  Pejcha}]{poddany_exoplanet_2010}
Poddaný, S., Brát, L., \& Pejcha, O. 2010, New Astronomy, 15, 297,
  \dodoi{10.1016/j.newast.2009.09.001}

\bibitem[{{P{\"u}sk{\"u}ll{\"u}} {et~al.}(2017){P{\"u}sk{\"u}ll{\"u}},
  {Soydugan}, {Erdem}, \& {Budding}}]{puskullu_photometric_2017}
{P{\"u}sk{\"u}ll{\"u}}, {\c{C}}., {Soydugan}, F., {Erdem}, A., \& {Budding}, E.
  2017, \na, 55, 39, \dodoi{10.1016/j.newast.2017.04.001}

\bibitem[{Rabus {et~al.}(2008)Rabus, Alonso, Deeg, Belmonte, Almenara,
  Gilliland, \& Brown}]{rabus_transit_2008}
Rabus, M., Alonso, R., Deeg, H.~J., {et~al.} 2008, Proceedings of the
  International Astronomical Union, 4, 432, \dodoi{10.1017/S1743921308026859}

\bibitem[{Rabus {et~al.}(2009)Rabus, Deeg, Alonso, Belmonte, \&
  Almenara}]{rabus_transit_2009}
Rabus, M., Deeg, H.~J., Alonso, R., Belmonte, J.~A., \& Almenara, J.~M. 2009,
  Astronomy and Astrophysics, 508, 1011, \dodoi{10.1051/0004-6361/200912252}

\bibitem[{Raetz {et~al.}(2014)Raetz, Maciejewski, Ginski, Mugrauer, Berndt,
  Eisenbeiss, Adam, Raetz, Roell, Seeliger, Marka, Vaňko, Bukowiecki, Errmann,
  Kitze, Ohlert, Pribulla, Schmidt, Sebastian, Puchalski, Tetzlaff, Hohle,
  Schmidt, \& Neuhäuser}]{raetz_transit_2014}
Raetz, S., Maciejewski, G., Ginski, C., {et~al.} 2014, MNRAS, 444, 1351,
  \dodoi{10.1093/mnras/stu1505}

\bibitem[{Ragozzine \& Wolf(2009)}]{ragozzine_probing_2009}
Ragozzine, D., \& Wolf, A.~S. 2009, The Astrophysical Journal, 698, 1778,
  \dodoi{10.1088/0004-637X/698/2/1778}

\bibitem[{Ricci {et~al.}(2014)Ricci, Ramón-Fox, Ayala-Loera, Michel,
  Navarro-Meza, Fox-Machado, Reyes-Ruiz, Sevilla, \&
  Curiel}]{ricci_multi-filter_2014}
Ricci, D., Ramón-Fox, F.~G., Ayala-Loera, C., {et~al.} 2014,
  \dodoi{10.1086/680233}

\bibitem[{Ricker {et~al.}(2015)Ricker, Winn, Vanderspek, Latham, Bakos, Bean,
  Berta-Thompson, Brown, Buchhave, Butler, Butler, Chaplin, Charbonneau,
  Christensen-Dalsgaard, Clampin, Deming, Doty, De~Lee, Dressing, Dunham, Endl,
  Fressin, Ge, Henning, Holman, Howard, Ida, Jenkins, Jernigan, Johnson,
  Kaltenegger, Kawai, Kjeldsen, Laughlin, Levine, Lin, Lissauer, MacQueen,
  Marcy, McCullough, Morton, Narita, Paegert, Palle, Pepe, Pepper, Quirrenbach,
  Rinehart, Sasselov, Sato, Seager, Sozzetti, Stassun, Sullivan, Szentgyorgyi,
  Torres, Udry, \& Villasenor}]{ricker_transiting_2015}
Ricker, G.~R., Winn, J.~N., Vanderspek, R., {et~al.} 2015, Journal of
  astronomical telescopes, instruments, and systems, 1, 014003,
  \dodoi{10.1117/1.JATIS.1.1.014003}

\bibitem[{Schröter {et~al.}(2012)Schröter, Schmitt, \&
  Müller}]{schroter_consistent_2012}
Schröter, S., Schmitt, J. H. M.~M., \& Müller, H.~M. 2012, Astronomy and
  Astrophysics, 539, A97, \dodoi{10.1051/0004-6361/201118536}

\bibitem[{Seeliger {et~al.}(2014)Seeliger, Dimitrov, Kjurkchieva, Mallonn,
  Fernandez, Kitze, Casanova, Maciejewski, Ohlert, Schmidt, Pannicke,
  Puchalski, Göğüş, Güver, Bilir, Ak, Hohle, Schmidt, Errmann, Jensen,
  Cohen, Marschall, Saral, Bernt, Derman, Gałan, \&
  Neuhäuser}]{seeliger_transit_2014}
Seeliger, M., Dimitrov, D., Kjurkchieva, D., {et~al.} 2014, MNRAS, 441, 304,
  \dodoi{10.1093/mnras/stu567}

\bibitem[{Seeliger {et~al.}(2015)Seeliger, Kitze, Errmann, Richter, Ohlert,
  Chen, Guo, Göğüş, Güver, Aydın, Mottola, Hellmich, Fernandez, Aceituno,
  Dimitrov, Kjurkchieva, Jensen, Cohen, Kundra, Pribulla, Vaňko, Budaj,
  Mallonn, Wu, Zhou, Raetz, Adam, Schmidt, Ide, Mugrauer, Marschall, Hackstein,
  Chini, Haas, Ak, Güzel, Özdönmez, Ginski, Marka, Schmidt, Dincel, Werner,
  Dathe, Greif, Wolf, Buder, Pannicke, Puchalski, \&
  Neuhäuser}]{seeliger_ground-based_2015}
Seeliger, M., Kitze, M., Errmann, R., {et~al.} 2015, MNRAS, 451, 4060,
  \dodoi{10.1093/mnras/stv1187}

\bibitem[{Sokov {et~al.}(2018)Sokov, Sokova, Dyachenko, Rastegaev, Burdanov,
  Rusov, Benni, Shadick, Hentunen, Salisbury, Esseiva, Garlitz, Bretton, Ogmen,
  Karavaev, Ayiomamitis, Mazurenko, Alonso, \& Velichko}]{sokov_transit_2018}
Sokov, E.~N., Sokova, I.~A., Dyachenko, V.~V., {et~al.} 2018, MNRAS, 480, 291,
  \dodoi{10.1093/mnras/sty1615}

\bibitem[{Sonbas {et~al.}(2021)Sonbas, Karaman, Özdönmez, Er, Dhuga,
  Göğüş, Nasiroglu, \& Zejmo}]{sonbas_probing_2021}
Sonbas, E., Karaman, N., Özdönmez, A., {et~al.} 2021, MNRAS, 509, 5102,
  \dodoi{10.1093/mnras/stab3270}

\bibitem[{{Sousa-Silva} {et~al.}(2018){Sousa-Silva}, {McKemmish}, {Chubb},
  {Gorman}, {Baker}, {Barton}, {Rivlin}, \& {Tennyson}}]{sousa_silva_orbyts}
{Sousa-Silva}, C., {McKemmish}, L.~K., {Chubb}, K.~L., {et~al.} 2018, Physics
  Education, 53, 015020, \dodoi{10.1088/1361-6552/aa8f2a}

\bibitem[{Southworth {et~al.}(2019)Southworth, Dominik, Jørgensen, Andersen,
  Bozza, Burgdorf, D’Ago, Dib, Figuera Jaimes, Fujii, Gill, Haikala, Hinse,
  Hundertmark, Khalouei, Korhonen, Longa-Peña, Mancini, Peixinho, Rabus,
  Rahvar, Sajadian, Skottfelt, Snodgrass, Spyratos, Tregloan-Reed,
  Unda-Sanzana, \& von Essen}]{southworth_transit_2019}
Southworth, J., Dominik, M., Jørgensen, U.~G., {et~al.} 2019, MNRAS, 490,
  4230, \dodoi{10.1093/mnras/stz2602}

\bibitem[{Sozzetti {et~al.}(2009)Sozzetti, Torres, Charbonneau, Winn,
  Korzennik, Holman, Latham, Laird, Fernandez, O'Donovan, Mandushev, Dunham,
  Everett, Esquerdo, Rabus, Belmonte, Deeg, Brown, Hidas, \&
  Baliber}]{sozzetti_new_2009}
Sozzetti, A., Torres, G., Charbonneau, D., {et~al.} 2009, The Astrophysical
  Journal, 691, 1145, \dodoi{10.1088/0004-637X/691/2/1145}

\bibitem[{{Tinetti} {et~al.}(2018){Tinetti}, {Drossart}, {Eccleston},
  {Hartogh}, {Heske}, {Leconte}, {Micela}, {Ollivier}, {Pilbratt}, {Puig},
  {Turrini}, {Vandenbussche}, {Wolkenberg}, {Beaulieu}, {Buchave}, {Ferus},
  {Griffin}, {Guedel}, {Justtanont}, {Lagage}, {Machado}, {Malaguti}, {Min},
  {N{\o}rgaard-Nielsen}, {Rataj}, {Ray}, {Ribas}, {Swain}, {Szabo}, {Werner},
  {Barstow}, {Burleigh}, {Cho}, {du Foresto}, {Coustenis}, {Decin}, {Encrenaz},
  {Galand}, {Gillon}, {Helled}, {Morales}, {Mu{\~n}oz}, {Moneti}, {Pagano},
  {Pascale}, {Piccioni}, {Pinfield}, {Sarkar}, {Selsis}, {Tennyson}, {Triaud},
  {Venot}, {Waldmann}, {Waltham}, {Wright}, {Amiaux}, {Augu{\`e}res},
  {Berth{\'e}}, {Bezawada}, {Bishop}, {Bowles}, {Coffey}, {Colom{\'e}},
  {Crook}, {Crouzet}, {Da Peppo}, {Sanz}, {Focardi}, {Frericks}, {Hunt},
  {Kohley}, {Middleton}, {Morgante}, {Ottensamer}, {Pace}, {Pearson},
  {Stamper}, {Symonds}, {Rengel}, {Renotte}, {Ade}, {Affer}, {Alard}, {Allard},
  {Altieri}, {Andr{\'e}}, {Arena}, {Argyriou}, {Aylward}, {Baccani}, {Bakos},
  {Banaszkiewicz}, {Barlow}, {Batista}, {Bellucci}, {Benatti}, {Bernardi},
  {B{\'e}zard}, {Blecka}, {Bolmont}, {Bonfond}, {Bonito}, {Bonomo}, {Brucato},
  {Brun}, {Bryson}, {Bujwan}, {Casewell}, {Charnay}, {Pestellini}, {Chen},
  {Ciaravella}, {Claudi}, {Cl{\'e}dassou}, {Damasso}, {Damiano}, {Danielski},
  {Deroo}, {Di Giorgio}, {Dominik}, {Doublier}, {Doyle}, {Doyon}, {Drummond},
  {Duong}, {Eales}, {Edwards}, {Farina}, {Flaccomio}, {Fletcher}, {Forget},
  {Fossey}, {Fr{\"a}nz}, {Fujii}, {Garc{\'\i}a-Piquer}, {Gear}, {Geoffray},
  {G{\'e}rard}, {Gesa}, {Gomez}, {Graczyk}, {Griffith}, {Grodent}, {Guarcello},
  {Gustin}, {Hamano}, {Hargrave}, {Hello}, {Heng}, {Herrero}, {Hornstrup},
  {Hubert}, {Ida}, {Ikoma}, {Iro}, {Irwin}, {Jarchow}, {Jaubert}, {Jones},
  {Julien}, {Kameda}, {Kerschbaum}, {Kervella}, {Koskinen}, {Krijger}, {Krupp},
  {Lafarga}, {Landini}, {Lellouch}, {Leto}, {Luntzer}, {Rank-L{\"u}ftinger},
  {Maggio}, {Maldonado}, {Maillard}, {Mall}, {Marquette}, {Mathis}, {Maxted},
  {Matsuo}, {Medvedev}, {Miguel}, {Minier}, {Morello}, {Mura}, {Narita},
  {Nascimbeni}, {Nguyen Tong}, {Noce}, {Oliva}, {Palle}, {Palmer}, {Pancrazzi},
  {Papageorgiou}, {Parmentier}, {Perger}, {Petralia}, {Pezzuto},
  {Pierrehumbert}, {Pillitteri}, {Piotto}, {Pisano}, {Prisinzano}, {Radioti},
  {R{\'e}ess}, {Rezac}, {Rocchetto}, {Rosich}, {Sanna}, {Santerne}, {Savini},
  {Scandariato}, {Sicardy}, {Sierra}, {Sindoni}, {Skup}, {Snellen}, {Sobiecki},
  {Soret}, {Sozzetti}, {Stiepen}, {Strugarek}, {Taylor}, {Taylor}, {Terenzi},
  {Tessenyi}, {Tsiaras}, {Tucker}, {Valencia}, {Vasisht}, {Vazan}, {Vilardell},
  {Vinatier}, {Viti}, {Waters}, {Wawer}, {Wawrzaszek}, {Whitworth}, {Yung},
  {Yurchenko}, {Osorio}, {Zellem}, {Zingales}, \& {Zwart}}]{tinetti_ariel}
{Tinetti}, G., {Drossart}, P., {Eccleston}, P., {et~al.} 2018, Experimental
  Astronomy, 46, 135, \dodoi{10.1007/s10686-018-9598-x}

\bibitem[{{Tinetti} {et~al.}(2021){Tinetti}, {Eccleston}, {Haswell}, {Lagage},
  {Leconte}, {L{\"u}ftinger}, {Micela}, {Min}, {Pilbratt}, {Puig}, {Swain},
  {Testi}, {Turrini}, {Vandenbussche}, {Rosa Zapatero Osorio}, {Aret},
  {Beaulieu}, {Buchhave}, {Ferus}, {Griffin}, {Guedel}, {Hartogh}, {Machado},
  {Malaguti}, {Pall{\'e}}, {Rataj}, {Ray}, {Ribas}, {Szab{\'o}}, {Tan},
  {Werner}, {Ratti}, {Scharmberg}, {Salvignol}, {Boudin}, {Halain}, {Haag},
  {Crouzet}, {Kohley}, {Symonds}, {Renk}, {Caldwell}, {Abreu}, {Alonso},
  {Amiaux}, {Berth{\'e}}, {Bishop}, {Bowles}, {Carmona}, {Coffey},
  {Colom{\'e}}, {Crook}, {D{\'e}sjonqueres}, {D{\'\i}az}, {Drummond},
  {Focardi}, {G{\'o}mez}, {Holmes}, {Krijger}, {Kovacs}, {Hunt}, {Machado},
  {Morgante}, {Ollivier}, {Ottensamer}, {Pace}, {Pagano}, {Pascale}, {Pearson},
  {M{\o}ller Pedersen}, {Pniel}, {Roose}, {Savini}, {Stamper}, {Szirovicza},
  {Szoke}, {Tosh}, {Vilardell}, {Barstow}, {Borsato}, {Casewell}, {Changeat},
  {Charnay}, {Civi{\v{s}}}, {Coud{\'e} du Foresto}, {Coustenis}, {Cowan},
  {Danielski}, {Demangeon}, {Drossart}, {Edwards}, {Gilli}, {Encrenaz}, {Kiss},
  {Kokori}, {Ikoma}, {Morales}, {Mendon{\c{c}}a}, {Moneti}, {Mugnai},
  {Garc{\'\i}a Mu{\~n}oz}, {Helled}, {Kama}, {Miguel}, {Nikolaou}, {Pagano},
  {Panic}, {Rengel}, {Rickman}, {Rocchetto}, {Sarkar}, {Selsis}, {Tennyson},
  {Tsiaras}, {Venot}, {Vida}, {Waldmann}, {Yurchenko}, {Szab{\'o}}, {Zellem},
  {Al-Refaie}, {Perez Alvarez}, {Anisman}, {Arhancet}, {Ateca}, {Baeyens},
  {Barnes}, {Bell}, {Benatti}, {Biazzo}, {B{\l}{\k{e}}cka}, {Bonomo}, {Bosch},
  {Bossini}, {Bourgalais}, {Brienza}, {Brucalassi}, {Bruno}, {Caines},
  {Calcutt}, {Campante}, {Canestrari}, {Cann}, {Casali}, {Casas}, {Cassone},
  {Cara}, {Carmona}, {Carone}, {Carrasco}, {Changeat}, {Chioetto},
  {Cortecchia}, {Czupalla}, {Chubb}, {Ciaravella}, {Claret}, {Claudi},
  {Codella}, {Garcia Comas}, {Cracchiolo}, {Cubillos}, {Da Peppo}, {Decin},
  {Dejabrun}, {Delgado-Mena}, {Di Giorgio}, {Diolaiti}, {Dorn}, {Doublier},
  {Doumayrou}, {Dransfield}, {Dumaye}, {Dunford}, {Jimenez Escobar}, {Van
  Eylen}, {Farina}, {Fedele}, {Fern{\'a}ndez}, {Fleury}, {Fonte}, {Fontignie},
  {Fossati}, {Funke}, {Galy}, {Garai}, {Garc{\'\i}a}, {Garc{\'\i}a-Rigo},
  {Garufi}, {Germano Sacco}, {Giacobbe}, {G{\'o}mez}, {Gonzalez},
  {Gonzalez-Galindo}, {Grassi}, {Griffith}, {Guarcello}, {Goujon}, {Gressier},
  {Grzegorczyk}, {Guillot}, {Guilluy}, {Hargrave}, {Hellin}, {Herrero},
  {Hills}, {Horeau}, {Ito}, {Jessen}, {Kabath}, {K{\'a}lm{\'a}n}, {Kawashima},
  {Kimura}, {Kn{\'\i}{\v{z}}ek}, {Kreidberg}, {Kruid}, {Kruijssen},
  {Kubel{\'\i}k}, {Lara}, {Lebonnois}, {Lee}, {Lefevre}, {Lichtenberg},
  {Locci}, {Lombini}, {Sanchez Lopez}, {Lorenzani}, {MacDonald}, {Magrini},
  {Maldonado}, {Marcq}, {Migliorini}, {Modirrousta-Galian}, {Molaverdikhani},
  {Molinari}, {Molli{\`e}re}, {Moreau}, {Morello}, {Morinaud}, {Morvan},
  {Moses}, {Mouzali}, {Nakhjiri}, {Naponiello}, {Narita}, {Nascimbeni},
  {Nikolaou}, {Noce}, {Oliva}, {Palladino}, {Papageorgiou}, {Parmentier},
  {Peres}, {P{\'e}rez}, {Perez-Hoyos}, {Perger}, {Cecchi Pestellini},
  {Petralia}, {Philippon}, {Piccialli}, {Pignatari}, {Piotto}, {Podio},
  {Polenta}, {Preti}, {Pribulla}, {Lopez Puertas}, {Rainer}, {Reess}, {Rimmer},
  {Robert}, {Rosich}, {Rossi}, {Rust}, {Saleh}, {Sanna}, {Schisano},
  {Schreiber}, {Schwartz}, {Scippa}, {Seli}, {Shibata}, {Simpson}, {Shorttle},
  {Skaf}, {Skup}, {Sobiecki}, {Sousa}, {Sozzetti}, {{\v{S}}poner}, {Steiger},
  {Tanga}, {Tackley}, {Taylor}, {Tecza}, {Terenzi}, {Tremblin}, {Tozzi},
  {Triaud}, {Trompet}, {Tsai}, {Tsantaki}, {Valencia}, {Carine Vandaele}, {Van
  der Swaelmen}, {Vardan}, {Vasisht}, {Vazan}, {Del Vecchio}, {Waltham},
  {Wawer}, {Widemann}, {Wolkenberg}, {Hou Yip}, {Yung}, {Zilinskas},
  {Zingales}, \& {Zuppella}}]{tinetti_ariel2}
{Tinetti}, G., {Eccleston}, P., {Haswell}, C., {et~al.} 2021, arXiv e-prints,
  arXiv:2104.04824.
\newblock \doarXiv{2104.04824}

\bibitem[{Turner {et~al.}(2022)Turner, Flagg, Ridden-Harper, \&
  Jayawardhana}]{turner_characterizing_2022}
Turner, J.~D., Flagg, L., Ridden-Harper, A., \& Jayawardhana, R. 2022, The
  Astronomical Journal, 163, \dodoi{10.3847/1538-3881/ac686f}

\bibitem[{Turner {et~al.}(2020)Turner, Ridden-Harper, \&
  Jayawardhana}]{turner_decaying_2020}
Turner, J.~D., Ridden-Harper, A., \& Jayawardhana, R. 2020,
  \dodoi{10.3847/1538-3881/abd178}

\bibitem[{Vanko {et~al.}(2013)Vanko, Maciejewski, Jakubík, Krejcová, Budaj,
  Pribulla, Ohlert, Raetz, Parimucha, \& Bukowiecki}]{vanko_photometric_2013}
Vanko, M., Maciejewski, G., Jakubík, M., {et~al.} 2013, MNRAS, 432, 944,
  \dodoi{10.1093/mnras/stt502}

\bibitem[{Wilson {et~al.}(2008)Wilson, Gillon, Hellier, Maxted, Pepe, Queloz,
  Anderson, Cameron, Smalley, Lister, Bentley, Blecha, Christian, Enoch,
  Haswell, Hebb, Horne, Irwin, Joshi, Kane, Marmier, Mayor, Parley, Pollacco,
  Pont, Ryans, Segransan, Skillen, Street, Udry, West, \&
  Wheatley}]{wilson_wasp-4b_2008}
Wilson, D.~M., Gillon, M., Hellier, C., {et~al.} 2008, The Astrophysical
  Journal, 675, L113, \dodoi{10.1086/586735}

\bibitem[{Wong {et~al.}(2022)Wong, Shporer, Vissapragada, Greklek-McKeon,
  Knutson, Winn, \& Benneke}]{wong_tess_2022}
Wong, I., Shporer, A., Vissapragada, S., {et~al.} 2022, The Astronomical
  Journal, 163, \dodoi{10.3847/1538-3881/ac5680}

\bibitem[{Yee {et~al.}(2019)Yee, Winn, Knutson, Patra, Vissapragada, Zhang,
  Holman, Shporer, \& Wright}]{yee_orbit_2019}
Yee, S.~W., Winn, J.~N., Knutson, H.~A., {et~al.} 2019, Astrophysical journal.
  Letters, 888, L5, \dodoi{10.3847/2041-8213/ab5c16}

\bibitem[{{Zellem} {et~al.}(2020){Zellem}, {Pearson}, {Blaser}, {Fowler},
  {Ciardi}, {Biferno}, {Massey}, {Marchis}, {Baer}, {Ball}, {Chasin}, {Conley},
  {Dixon}, {Fletcher}, {Hernandez}, {Nair}, {Perian}, {Sienkiewicz}, {Tock},
  {Vijayakumar}, {Swain}, {Roudier}, {Bryden}, {Conti}, {Hill}, {Hergenrother},
  {Dussault}, {Kane}, {Fitzgerald}, {Boyce}, {Peticolas}, {Gee}, {Cominsky},
  {Zimmerman-Brachman}, {Smith}, {Creech-Eakman}, {Engelke}, {Iturralde},
  {Dragomir}, {Jovanovic}, {Lawton}, {Arbouch}, {Kuchner}, \&
  {Malvache}}]{zellem_2020}
{Zellem}, R.~T., {Pearson}, K.~A., {Blaser}, E., {et~al.} 2020, \pasp, 132,
  054401, \dodoi{10.1088/1538-3873/ab7ee7}

\bibitem[{Zhao {et~al.}(2018)Zhao, Jiang-hui, \& Yao}]{zhao_photometric_2018}
Zhao, S., Jiang-hui, J., \& Yao, D. 2018, Chinese Astronomy and Astrophysics,
  42, 101, \dodoi{10.1016/j.chinastron.2018.01.007}

\end{thebibliography}
\bibliographystyle{aasjournal}

%%% APPENDICES %%%
\newpage
\restartappendixnumbering
\appendix
\section{Full MCMC Results} \label{sec:appendixA}

\vspace{5mm}
The results of the Metropolis-Hastings MCMC sampling of the constant period, orbital decay, and apsidal precession transit timing models are presented below. The top ten targets of interest, as discussed in section \ref{sec:results}, are listed alphabetically. 
\vspace{5mm}

\startlongtable
\begin{deluxetable*}{lcccc}
\tablecolumns{5}
\tablewidth{0pt}
\tablecaption{HAT-P-19\,b Results}
\tablehead{\colhead{Parameter} & \colhead{Symbol} & \colhead{Unit} & \colhead{Value} & \colhead{Uncertainty (1-$\sigma$)}}
\startdata
\sidehead{\textbf{Constant Period Model}}
Transit Center & $t_{0}$ & BJD$_{\rm TDB}$ & $2455091.53464$ & $^{+0.00010}_{-0.00010}$ \vspace{0.1cm} \\ 
Period & $P$ & days & $4.00878388$ & $^{+0.00000015}_{-0.00000014}$ \vspace{0.1cm} \\
Number of Data & $n$ & & $72$ \vspace{0.1cm} \\ 
Degrees of Freedom & $k$ & & $2$ \vspace{0.1cm} \\ 
Chi-Square Statistic & $\chi^2$ & & $261.8$ \vspace{0.1cm} \\
Bayesian Information Criterion & BIC & & $270.4$ \vspace{0.1cm} \\ 
\sidehead{\textbf{Orbital Decay Model}}
Transit Center & $t_{0}$ & BJD$_{\rm TDB}$ & $2455091.53382$ & $^{+0.00014}_{-0.00015}$ \vspace{0.1cm} \\
Period & $P$ & days & $4.00878805$ & $^{+0.00000055}_{-0.00000054}$ \vspace{0.1cm} \\
\enddata
\vspace{0.02cm}
\label{table:hatp19results}
\end{deluxetable*} 
\startlongtable
\begin{deluxetable*}{lcccc}
\tablecolumns{5}
\tablewidth{0pt}
\tablecaption{HAT-P-32\,b Results}
\tablehead{\colhead{Parameter} & \colhead{Symbol} & \colhead{Unit} & \colhead{Value} & \colhead{Uncertainty (1-$\sigma$)}}
\startdata
\sidehead{\textbf{Constant Period Model}}
Transit Center & $t_{0}$ & BJD$_{\rm TDB}$ & $2454420.447196$ & $^{+0.000052}_{-0.000051}$ \vspace{0.1cm} \\ 
Period & $P$ & days & $2.150008203$ & $^{+0.000000038}_{-0.000000038}$ \vspace{0.1cm} \\
Number of Data & $n$ & & $87$ \vspace{0.1cm} \\ 
Degrees of Freedom & $k$ & & $2$ \vspace{0.1cm} \\ 
Chi-Square Statistic & $\chi^2$ & & $668.3$ \vspace{0.1cm} \\
Bayesian Information Criterion & BIC & & $677.2$ \vspace{0.1cm} \\ 
\sidehead{\textbf{Orbital Decay Model}}
Transit Center & $t_{0}$ & BJD$_{\rm TDB}$ & $2454420.447035$ & $^{+0.000061}_{-0.000061}$ \vspace{0.1cm} \\
Period & $P$ & days & $2.15000874$ & $^{+0.00000012}_{-0.00000012}$ \vspace{0.1cm} \\
\enddata
\vspace{0.02cm}
\label{table:hatp32results}
\end{deluxetable*} \newpage
\startlongtable
\begin{deluxetable*}{lcccc}
\tablecolumns{5}
\tablewidth{0pt}
\tablecaption{TrES-1\,b Results}
\tablehead{\colhead{Parameter} & \colhead{Symbol} & \colhead{Unit} & \colhead{Value} & \colhead{Uncertainty (1-$\sigma$)}}
\startdata
\sidehead{\textbf{Constant Period Model}}
Transit Center & $t_{0}$ & BJD$_{\rm TDB}$ & $2453898.874169$ & $^{+0.000038}_{-0.000038}$ \vspace{0.1cm} \\ 
Period & $P$ & days & $3.030069676$ & $^{+0.000000042}_{-0.000000042}$ \vspace{0.1cm} \\
Number of Data & $n$ & & $66$ \vspace{0.1cm} \\ 
Degrees of Freedom & $k$ & & $2$ \vspace{0.1cm} \\ 
Chi-Square Statistic & $\chi^2$ & & $182.6$ \vspace{0.1cm} \\
Bayesian Information Criterion & BIC & & $190.9$ \vspace{0.1cm} \\ 
\sidehead{\textbf{Orbital Decay Model}}
Transit Center & $t_{0}$ & BJD$_{\rm TDB}$ & $2453898.874084$ & $^{+0.000041}_{-0.000041}$ \vspace{0.1cm} \\
Period & $P$ & days & $3.03007055$ & $^{+0.00000017}_{-0.00000017}$ \vspace{0.1cm} \\
\enddata
\vspace{0.02cm}
\label{table:tres1results}
\end{deluxetable*} \newpage
\startlongtable
\begin{deluxetable*}{lcccc}
\tablecolumns{5}
\tablewidth{0pt}
\tablecaption{TrES-2\,b Results}
\tablehead{\colhead{Parameter} & \colhead{Symbol} & \colhead{Unit} & \colhead{Value} & \colhead{Uncertainty (1-$\sigma$)}}
\startdata
\sidehead{\textbf{Constant Period Model}}
Transit Center & $t_{0}$ & BJD$_{\rm TDB}$ & $2453957.635474$ & $^{+0.000020}_{-0.000020}$ \vspace{0.1cm} \\ 
Period & $P$ & days & $2.47061340$ & $^{+0.000000042}_{-0.000000042}$ \vspace{0.1cm} \\
Number of Data & $n$ & & $149$ \vspace{0.1cm} \\ 
Degrees of Freedom & $k$ & & $2$ \vspace{0.1cm} \\ 
Chi-Square Statistic & $\chi^2$ & & $868.8$ \vspace{0.1cm} \\
Bayesian Information Criterion & BIC & & $878.9$ \vspace{0.1cm} \\ 
\sidehead{\textbf{Orbital Decay Model}}
Transit Center & $t_{0}$ & BJD$_{\rm TDB}$ & $2453957.635119$ & $^{+0.000069}_{-0.000070}$ \vspace{0.1cm} \\
Period & $P$ & days & $2.47061447$ & $^{+0.00000020}_{-0.00000020}$ \vspace{0.1cm} \\
\enddata
\vspace{0.02cm}
\label{table:tres2results}
\end{deluxetable*} \newpage
\startlongtable
\begin{deluxetable*}{lcccc}
\tablecolumns{5}
\tablewidth{0pt}
\tablecaption{TrES-3\,b Results}
\tablehead{\colhead{Parameter} & \colhead{Symbol} & \colhead{Unit} & \colhead{Value} & \colhead{Uncertainty (1-$\sigma$)}}
\startdata
\sidehead{\textbf{Constant Period Model}}
Transit Center & $t_{0}$ & BJD$_{\rm TDB}$ & $2454538.581479$ & $^{+0.000027}_{-0.000027}$ \vspace{0.1cm} \\ 
Period & $P$ & days & $1.306186320$ & $^{+0.000000014}_{-0.000000014}$ \vspace{0.1cm} \\
Number of Data & $n$ & & $218$ \vspace{0.1cm} \\ 
Degrees of Freedom & $k$ & & $2$ \vspace{0.1cm} \\ 
Chi-Square Statistic & $\chi^2$ & & $929.0$ \vspace{0.1cm} \\
Bayesian Information Criterion & BIC & & $939.8$ \vspace{0.1cm} \\ 
\sidehead{\textbf{Orbital Decay Model}}
Transit Center & $t_{0}$ & BJD$_{\rm TDB}$ & $2454538.581437$ & $^{+0.000030}_{-0.000030}$ \vspace{0.1cm} \\
Period & $P$ & days & $1.306186497$ & $^{+0.000000050}_{-0.000000052}$ \vspace{0.1cm} \\
\enddata
\vspace{0.02cm}
\label{table:tres3results}
\end{deluxetable*} \newpage
\startlongtable
\begin{deluxetable*}{lcccc}
\tablecolumns{5}
\tablewidth{0pt}
\tablecaption{TrES-5\,b Results}
\tablehead{\colhead{Parameter} & \colhead{Symbol} & \colhead{Unit} & \colhead{Value} & \colhead{Uncertainty (1-$\sigma$)}}
\startdata
\sidehead{\textbf{Constant Period Model}}
Transit Center & $t_{0}$ & BJD$_{\rm TDB}$ & $2455443.25340$ & $^{+0.00012}_{-0.00012}$ \vspace{0.1cm} \\ 
Period & $P$ & days & $1.482246663$ & $^{+0.000000064}_{-0.000000064}$ \vspace{0.1cm} \\
Number of Data & $n$ & & $109$ \vspace{0.1cm} \\ 
Degrees of Freedom & $k$ & & $2$ \vspace{0.1cm} \\ 
Chi-Square Statistic & $\chi^2$ & & $531.8$ \vspace{0.1cm} \\
Bayesian Information Criterion & BIC & & $541.1$ \vspace{0.1cm} \\ 
\sidehead{\textbf{Orbital Decay Model}}
Transit Center & $t_{0}$ & BJD$_{\rm TDB}$ & $2455443.25109$ & $^{+0.00033}_{-0.00032}$ \vspace{0.1cm} \\
Period & $P$ & days & $1.48224954$ & $^{+0.00000038}_{-0.00000039}$ \vspace{0.1cm} \\
\enddata
\vspace{0.02cm}
\label{table:tres5results}
\end{deluxetable*} \newpage
\startlongtable
\begin{deluxetable*}{lcccc}
\tablecolumns{5}
\tablewidth{0pt}
\tablecaption{WASP-4\,b Results}
\tablehead{\colhead{Parameter} & \colhead{Symbol} & \colhead{Unit} & \colhead{Value} & \colhead{Uncertainty (1-$\sigma$)}}
\startdata
\sidehead{\textbf{Constant Period Model}}
Transit Center & $t_{0}$ & BJD$_{\rm TDB}$ & $2455880.794607$ & $^{+0.000018}_{-0.000018}$ \vspace{0.1cm} \\ 
Period & $P$ & days & $1.338231560$ & $^{+0.000000016}_{-0.000000015}$ \vspace{0.1cm} \\
Number of Data & $n$ & & $55$ \vspace{0.1cm} \\ 
Degrees of Freedom & $k$ & & $2$ \vspace{0.1cm} \\ 
Chi-Square Statistic & $\chi^2$ & & $396.9$ \vspace{0.1cm} \\
Bayesian Information Criterion & BIC & & $404.9$ \vspace{0.1cm} \\ 
\sidehead{\textbf{Orbital Decay Model}}
Transit Center & $t_{0}$ & BJD$_{\rm TDB}$ & $2455880.794814$ & $^{+0.000029}_{-0.000029}$ \vspace{0.1cm} \\
Period & $P$ & days & $1.338231695$ & $^{+0.000000022}_{-0.000000022}$ \vspace{0.1cm} \\
\enddata
\vspace{0.02cm}
\label{table:wasp4results}
\end{deluxetable*} \newpage
\startlongtable
\begin{deluxetable*}{lcccc}
\tablecolumns{5}
\tablewidth{0pt}
\tablecaption{WASP-10\,b Results}
\tablehead{\colhead{Parameter} & \colhead{Symbol} & \colhead{Unit} & \colhead{Value} & \colhead{Uncertainty (1-$\sigma$)}}
\startdata
\sidehead{\textbf{Constant Period Model}}
Transit Center & $t_{0}$ & BJD$_{\rm TDB}$ & $2454357.857968$ & $^{+0.000031}_{-0.000031}$ \vspace{0.1cm} \\ 
Period & $P$ & days & $3.092728113$ & $^{+0.000000044}_{-0.000000044}$ \vspace{0.1cm} \\
Number of Data & $n$ & & $122$ \vspace{0.1cm} \\ 
Degrees of Freedom & $k$ & & $2$ \vspace{0.1cm} \\ 
Chi-Square Statistic & $\chi^2$ & & $1001.7$ \vspace{0.1cm} \\
Bayesian Information Criterion & BIC & & $1011.3$ \vspace{0.1cm} \\ 
\sidehead{\textbf{Orbital Decay Model}}
Transit Center & $t_{0}$ & BJD$_{\rm TDB}$ & $2454357.857571$ & $^{+0.000054}_{-0.000054}$ \vspace{0.1cm} \\
Period & $P$ & days & $3.09272986$ & $^{+0.00000020}_{-0.00000020}$ \vspace{0.1cm} \\
\enddata
\vspace{0.02cm}
\label{table:wasp10results}
\end{deluxetable*} \newpage
\startlongtable
\begin{deluxetable*}{lcccc}
\tablecolumns{5}
\tablewidth{0pt}
\tablecaption{WASP-12\,b Results}
\tablehead{\colhead{Parameter} & \colhead{Symbol} & \colhead{Unit} & \colhead{Value} & \colhead{Uncertainty (1-$\sigma$)}}
\startdata
\sidehead{\textbf{Constant Period Model}}
Transit Center & $t_{0}$ & BJD$_{\rm TDB}$ & $2454508.978942$ & $^{+0.000046}_{-0.000047}$ \vspace{0.1cm} \\ 
Period & $P$ & days & $1.091419487$ & $^{+0.000000020}_{-0.000000020}$ \vspace{0.1cm} \\
Number of Data & $n$ & & $213$ \vspace{0.1cm} \\ 
Degrees of Freedom & $k$ & & $2$ \vspace{0.1cm} \\ 
Chi-Square Statistic & $\chi^2$ & & $2342.4$ \vspace{0.1cm} \\
Bayesian Information Criterion & BIC & & $2353.1$ \vspace{0.1cm} \\ 
\sidehead{\textbf{Orbital Decay Model}}
Transit Center & $t_{0}$ & BJD$_{\rm TDB}$ & $2454508.977198$ & $^{+0.000074}_{-0.000074}$ \vspace{0.1cm} \\
Period & $P$ & days & $1.091421946$ & $^{+0.000000084}_{-0.000000084}$ \vspace{0.1cm} \\
\enddata
\vspace{0.02cm}
\label{table:wasp12results}
\end{deluxetable*} \newpage
\startlongtable
\begin{deluxetable*}{lcccc}
\tablecolumns{5}
\tablewidth{0pt}
\tablecaption{WASP-43\,b Results}
\tablehead{\colhead{Parameter} & \colhead{Symbol} & \colhead{Unit} & \colhead{Value} & \colhead{Uncertainty (1-$\sigma$)}}
\startdata
\sidehead{\textbf{Constant Period Model}}
Transit Center & $t_{0}$ & BJD$_{\rm TDB}$ & $2455528.868421$ & $^{+0.000054}_{-0.000054}$ \vspace{0.1cm} \\ 
Period & $P$ & days & $0.813474243$ & $^{+0.000000017}_{-0.000000017}$ \vspace{0.1cm} \\
Number of Data & $n$ & & $116$ \vspace{0.1cm} \\ 
Degrees of Freedom & $k$ & & $2$ \vspace{0.1cm} \\ 
Chi-Square Statistic & $\chi^2$ & & $566.6$ \vspace{0.1cm} \\
Bayesian Information Criterion & BIC & & $576.2$ \vspace{0.1cm} \\ 
\sidehead{\textbf{Orbital Decay Model}}
Transit Center & $t_{0}$ & BJD$_{\rm TDB}$ & $2455528.868381$ & $^{+0.000072}_{-0.000072}$ \vspace{0.1cm} \\
Period & $P$ & days & $0.813474304$ & $^{+0.000000077}_{-0.000000072}$ \vspace{0.1cm} \\
\enddata
\label{table:wasp43results}
\end{deluxetable*} \newpage

\newpage
\section{Exoplanet Transit Database (ETD) Observations} \label{sec:appendixB}

\vspace{5mm}
The final cleaned data sets for the ten targets of interest (section \ref{sec:results}) are highlighted in the following tables to ensure reproducibility of these results. Observer names are also listed, as-given in the ETD with minor formatting changes. For entries where there are more than two observers, the full list of names is given in the table notes.
\vspace{5mm}

\startlongtable
\begin{deluxetable}{cllcl}
\tablecolumns{5}
\tablecaption{HAT-P-19\,b ETD Transit Centers}
\tablehead{\colhead{Epoch} & \colhead{Transit Center (BJD$_{\rm TDB}$)} & \colhead{Error} & \colhead{Data Quality} & \colhead{Observer}}
\startdata
0 & 2455091.534936 & 0.00034 & 1 &   Hartman et al. (2010) \\
97 & 2455480.388138842 & 0.00082 & 3 &   Ioannidis P., Avdellidou \\
97 & 2455480.388168842 & 0.00077 & 3 &  Lomoz F. \\
101 & 2455496.4200886358 & 0.00089 & 3 &  Naves R. \\
101 & 2455496.420158636 & 0.00072 & 3 &  Vanhuysse  M. \\
108 & 2455524.482118315 & 0.00036 & 1 &  Muler G. \\
108 & 2455524.482168315 & 0.00103 & 3 &  Husar D. \\
109 & 2455528.492328273 & 0.00043 & 2 &  Ruiz J. \\
\enddata
\tablecomments{\textsuperscript{*}Mikulecká B., Henych T. \textsuperscript{**}Qingfeng Pi, Aiying Zhou \textsuperscript{***}Ryan Cooney, Taylor Bell \textsuperscript{**†}Danilo Zardin, Marco Fiaschi}
\label{table:hatp19data}
\end{deluxetable}
\startlongtable
\begin{deluxetable}{cllcl}
\tablecolumns{5}
\tablecaption{HAT-P-32\,b ETD Transit Centers}
\tablehead{\colhead{Epoch} & \colhead{Transit Center (BJD$_{\rm TDB}$)} & \colhead{Error} & \colhead{Data Quality} & \colhead{Observer}}
\startdata
0 & 2454420.44712443 & 9e-05 & 1 &  Hartman et al. (2011) \\
640 & 2455796.449701029 & 0.001 & 3 &  Moudrá M., Sobotka P. \\
640 & 2455796.454151029 & 0.00085 & 3 &  Zibar M. \\
640 & 2455796.454951029 & 0.00058 & 2 &  Trnka J. \\
666 & 2455852.351390598 & 0.00072 & 2 &  Brosio A. \\
668 & 2455856.651900572 & 0.00026 & 1 &  Dax T. \\
668 & 2455856.6520805717 & 0.00029 & 1 &  Shadic S. \\
668 & 2455856.6526905717 & 0.00058 & 2 &  Yafunyaev M. \\
\enddata
\tablecomments{\textsuperscript{*}E. Helou, J. Lowenthal, R. O’Connor, E. Papineau, A. Peck, L. Stephens, K. Walker \textsuperscript{**}Johanna Hipp, Annalotta Hipp}
\label{table:hatp32data}
\end{deluxetable}
\startlongtable
\begin{deluxetable}{cllcl}
\tablecolumns{5}
\tablecaption{TrES-1\,b ETD Transit Centers}
\tablehead{\colhead{Epoch} & \colhead{Transit Center (BJD$_{\rm TDB}$)} & \colhead{Error} & \colhead{Data Quality} & \colhead{Observer}}
\startdata
-344 & 2452856.5293562952 & 0.0015 & 1 &    Charbonneau et al. (2005) \\ 
-340 & 2452868.6510561793 & 0.0022 & 1 &  Charbonneau et al. (2005) \\ 
-239 & 2453174.6871521757 & 0.0004 & 1 &  Charbonneau et al. (2005) \\ 
-236 & 2453183.775952023 & 0.0005 & 1 &  Charbonneau et al. (2005) \\ 
-235 & 2453186.806851972 & 0.0003 & 1 &    Charbonneau et al. (2005) \\ 
-234 & 2453189.8361519207 & 0.0019 & 1 &    Charbonneau et al. (2005) \\
-215 & 2453247.4082509303 & 0.0004 & 1 &  Charbonneau et al. (2005) \\ 
-208 & 2453268.6206305576 & 0.00073 & 3 &  Walsh B.\textsuperscript{*} \\
\enddata
\tablecomments{\textsuperscript{*}Balonek T., Iadanza K. \textsuperscript{**}Kuznietsova Y., Krushevska V. \textsuperscript{***}JT, MK, RD, TK, BH, MM, HK, JS (names unknown) \textsuperscript{**†}Strickland W., Soriano R. \textsuperscript{*††}Sevim K., Bastuk O. \textsuperscript{†††}Danilo Zardin, Marco Fiaschi}
\label{table:tres1data}
\end{deluxetable}
\startlongtable
\begin{deluxetable}{cllcl}
\tablecolumns{5}
\tablecaption{TrES-2\,b ETD Transit Centers}
\tablehead{\colhead{Epoch} & \colhead{Transit Center (BJD$_{\rm TDB}$)} & \colhead{Error} & \colhead{Data Quality} & \colhead{Observer}}
\startdata
0 & 2453957.6355417324 & 0.00038 & 1 &  Holman et al. (2007) \\ 
13 & 2453989.7536113295 & 0.00029 & 1 &  Holman et al. (2007) \\ 
15 & 2453994.694681268 & 0.00031 & 1 &  Holman et al. (2007) \\ 
34 & 2454041.6365406993 & 0.0003 & 1 &  Holman et al. (2007) \\ 
242 & 2454555.5269558984 & 0.00123 & 3 &  Brát L. \\ 
268 & 2454619.760645541 & 0.0013 & 2 &  Gary \\ 
281 & 2454651.876345383 & 0.0007 & 2 &  Moon \\ 
293 & 2454681.523145253 & 0.0021 & 3 &  Naves \\
\enddata
\tablecomments{\textsuperscript{*}Cihan Tugrul Tezcan, Ozgur Basturk \textsuperscript{**}Danilo Zardin, Marco Fiaschi}
\label{table:tres2data}
\end{deluxetable}
\startlongtable
\begin{deluxetable}{cllcl}
\tablecolumns{5}
\tablecaption{TrES-3\,b ETD Transit Centers}
\tablehead{\colhead{Epoch} & \colhead{Transit Center (BJD$_{\rm TDB}$)} & \colhead{Error} & \colhead{Data Quality} & \colhead{Observer}}
\startdata
-270 & 2454185.911171321 & 0.000198 & 1 &  Sozzeti et al. (2009) \\ 
-260 & 2454198.97388822 & 0.000223 & 1 &  Sozzetti et al. (2009) \\ 
-248 & 2454214.6470391033 & 0.00028 & 1 &  Sozzetti et al. (2009) \\ 
-247 & 2454215.952821094 & 0.000214 & 1 &  Sozzetti et al. (2009) \\ 
-229 & 2454239.462400929 & 0.00047 & 2 &  Fabjan T. et al. \\ 
-13 & 2454521.5984502416 & 0.00131 & 3 &  Brát L. \\ 
-3 & 2454534.663170264 & 0.00017 & 1 &  RISE telescope \\ 
-2 & 2454535.9689862663 & 0.000166 & 1 &  Sozzetti et al. (2009) \\
\enddata
\tablecomments{\textsuperscript{*}Stacy Irwin, and Karissa Haire. \textsuperscript{**}R. Campbell, D. Russell, and W. Strickland. \textsuperscript{***}M. Banfi, R. Papini, and G. Marino.}
\label{table:tres3data}
\end{deluxetable}
\startlongtable
\begin{deluxetable}{cllcl}
\tablecolumns{5}
\tablecaption{TrES-5\,b ETD Transit Centers}
\tablehead{\colhead{Epoch} & \colhead{Transit Center (BJD$_{\rm TDB}$)} & \colhead{Error} & \colhead{Data Quality} & \colhead{Observer}}
\startdata
544 & 2456249.595185279 & 0.00108 & 3 &  Shadic S. \\ 
558 & 2456270.3438554914 & 0.0006 & 2 &  Garcia F. \\ 
640 & 2456391.8897667313 & 0.00075 & 3 &  Shadic S. \\ 
646 & 2456400.78382682 & 0.00056 & 2 &  Shadic S. \\ 
671 & 2456437.8403671817 & 0.00102 & 3 &  Kehusmaa P., Harlingten C. \\ 
720 & 2456510.471477866 & 0.00086 & 3 &  Ayiomamitis A. \\ 
724 & 2456516.4012579215 & 0.00081 & 3 &  Benishek V. \\ 
741 & 2456541.5966481552 & 0.00089 & 3 &  Horta F. G. \\
\enddata
\label{table:tres5data}
\end{deluxetable}
\startlongtable
\begin{deluxetable}{cllcl}
\tablecolumns{5}
\tablecaption{WASP-4\,b ETD Transit Centers}
\tablehead{\colhead{Epoch} & \colhead{Transit Center (BJD$_{\rm TDB}$)} & \colhead{Error} & \colhead{Data Quality} & \colhead{Observer}}
\startdata
-1433 & 2453963.108629995 & 0.00081 & 1 &  Gillon et al. (2008) \\ 
-1133 & 2454364.5772199957 & 0.00075 & 1 &  Gillon et al. (2008) \\ 
-1132 & 2454365.9153884756 & 0.00025 & 3 &  Wilson et al. (2008) \\ 
-1130 & 2454368.5924399947 & 0.00022 & 1 &  Gillon et al. (2008) \\ 
-1128 & 2454371.2681200043 & 0.00033 & 1 &  Gillon et al. (2008) \\ 
-1109 & 2454396.6954099988 & 5.1e-05 & 1 &  Gillon et al. (2008) \\ 
-884 & 2454697.7981701 & 5.5e-05 & 1 &  Winn et al. (2009) \\ 
-846 & 2454748.6511103 & 7.2e-05 & 1 &  Winn et al. (2009) \\
\enddata
\tablecomments{ \textsuperscript{*}and Romina P. Di Sisto \textsuperscript{**}Hoňková K., Juryšek J.}
\label{table:wasp4data}
\end{deluxetable}
\startlongtable
\begin{deluxetable}{cllcl}
\tablecolumns{5}
\tablecaption{WASP-10\,b ETD Transit Centers}
\tablehead{\colhead{Epoch} & \colhead{Transit Center (BJD$_{\rm TDB}$)} & \colhead{Error} & \colhead{Data Quality} & \colhead{Observer}}
\startdata
0 & 2454357.85878443 & 0.00042 & 1 &  Christian et al. (2008) \\ 
99 & 2454664.0380455623 & 8.2e-05 & 1 &  John J. A. et al. (2008) \\ 
103 & 2454676.4088701825 & 0.00055 & 2 &  Brát L. \\ 
126 & 2454747.550048031 & 0.0006 & 3 &  Kalajian \\ 
135 & 2454775.378447218 & 0.0008 & 3 &  Krejcova T. et al. (2010) \\ 
138 & 2454784.6544469516 & 0.0004 & 2 &  Gary \\ 
139 & 2454787.7453468633 & 0.0007 & 2 &  Foote \\ 
155 & 2454837.231160002 & 0.00037 & 1 &  Maciejewski et al. (2010) \\
\enddata
\tablecomments{ \textsuperscript{*}M. Gaude, C. Gillier, J. Michelet, JP.Roux \textsuperscript{**}Lycée Notre Dame de Toutes Aides, France N. \textsuperscript{***}Ferreira, E. Helou, J. Lowenthal, R. O \textsuperscript{**†}M. Banfi, R. Papini, F. Salvaggio \textsuperscript{*††}B.A. Dumitru, D. Bertesteanu}
\label{table:wasp10data}
\end{deluxetable}
\startlongtable
\begin{deluxetable}{cllcl}
\tablecolumns{5}
\tablecaption{WASP-12\,b ETD Transit Centers}
\tablehead{\colhead{Epoch} & \colhead{Transit Center (BJD$_{\rm TDB}$)} & \colhead{Error} & \colhead{Data Quality} & \colhead{Observer}}
\startdata
0 & 2454508.97685446 & 0.0002 & 1 &  Hebb et al. (2008) \\ 
300 & 2454836.4033931005 & 0.0006 & 2 &  Hentunen \\ 
304 & 2454840.7686030436 & 0.00047 & 1 &  Tucker et al. (2011) \\ 
304 & 2454840.771193043 & 0.001 & 3 &  Gary \\ 
366 & 2454908.437992057 & 0.001 & 3 &  Naves \\ 
387 & 2454931.3581816694 & 0.00098 & 3 &  Brát L. \\ 
601 & 2455164.923956134 & 0.00149 & 3 &  Shadick S. \\ 
601 & 2455164.9243161334 & 0.00129 & 3 &  Stan Shadick \\
\enddata
\vspace{0.02cm}\tablecomments{ \textsuperscript{*}Danilo Zardin, Marco Fiaschi \textsuperscript{**}E. Helou, J. Lowenthal, R. O’Connor, E. Papineau, A. Peck, L. Stephens, K. Walker}
\label{table:wasp12data}
\end{deluxetable}
\startlongtable
\begin{deluxetable}{cllcl}
\tablecolumns{5}
\tablecaption{WASP-43\,b ETD Transit Centers}
\tablehead{\colhead{Epoch} & \colhead{Transit Center (BJD$_{\rm TDB}$)} & \colhead{Error} & \colhead{Data Quality} & \colhead{Observer}}
\startdata
0 & 2455528.8685292783 & 0.00014 & 1 &  Hellier et al. (2011) \\ 
497 & 2455933.16473013 & 0.00025 & 1 &  Starr P. \\ 
505 & 2455939.674749933 & 0.00052 & 2 &  Naves R. \\ 
506 & 2455940.487439908 & 0.0005 & 2 &  Ayiomamitis A. \\ 
511 & 2455944.554679785 & 0.00106 & 3 &  René R. \\ 
527 & 2455957.567029386 & 0.00156 & 3 &  František Lomoz \\ 
554 & 2455979.533498702 & 0.0004 & 2 &  Gonzalez J. \\ 
554 & 2455979.5339987027 & 0.00044 & 2 &  Nicolas E. \\
\enddata
\tablecomments{ \textsuperscript{*}Jacques Clement, Jean-Philippe Nougayrede \textsuperscript{**}Danilo Zardin, Marco Fiaschi \textsuperscript{***}Marana Space Explorer Center}
\label{table:wasp43data}
\end{deluxetable}

\end{document}